\documentclass[preprint]{aastex}

\slugcomment{To appear in Ap.J.}

\shorttitle{MASIV Survey. III}
\shortauthors{Pursimo, Ojha, Jauncey et al.}

\begin{document}


%\title{Redshifts and Optical Properties of Scintillating and Non-Scintillating
%Extragalactic Radio Sources in the MASIV survey}

\title{The Micro-Arcsecond Scintillation-Induced Variability 
(MASIV) Survey III. Optical Identifications and New Redshifts}


\author{Tapio Pursimo}
\affil{Nordic Optical Telescope, Apartado 474, 
	38700 Santa Cruz de La Palma, Spain}
\email{tpursimo@not.iac.es}
\author{Roopesh Ojha\altaffilmark{1}}
\affil{NVI Inc./ U. S. Naval Observatory, 3450 Massachusetts Ave NW, 
Washington DC, USA}
\altaffiltext{1}{Now at NASA/Goddard Space Flight Center/ORAU, Code 661, Astroparticle Physics Lab., Greenbelt, MD 20771, U.S.A.}

\author{David L. Jauncey}
%\affil{Australia Telescope National Facility CSIRO, PO Box 76, Epping, 
%NSW 1710, Australia}
\affil{CSIRO Astronomy and Space Science
and
Mount Stromlo Observatory,
Canberra ACT 0200
Australia}

\author{Barney J. Rickett}
\affil{Department of Electrical and Computer Engineering, University 
of California, San Diego, La Jolla, CA 92093, USA}

\author{Michael S. Dutka}
\affil{The Catholic University of America, 620 Michigan Ave., N.E., 
Washington DC 20064, U.S.A.}

\author{Jun Yi Koay\altaffilmark{2}}
\affil{ICRAR, Curtin University, Bentley, WA 6845, Australia}
\altaffiltext{2}{Now at Faculty of Engineering and
Science, Universiti Tunku Abdul Rahman, Setapak, KL 53300, Malaysia}

\author{James E. J. Lovell}
\affil{School of Mathematics and Physics, University of Tasmania, 
TAS 7001, Australia}

\author{Hayley E. Bignall}
\affil{ICRAR, Curtin University, Bentley, WA 6845, Australia}
\author{Lucyna Kedziora-Chudczer}
\affil{
School of Physics and Astrophysics, UNSW, Sydney NSW 2052
Australia}
\and
\author{Jean-Pierre Macquart}
\affil{ICRAR, Curtin University, Bentley, WA 6845, Australia}
 
\begin{abstract}

Intraday variability (IDV) of the radio emission from active galactic
nuclei is now known to be predominantly due to interstellar
scintillation (ISS). The MASIV (The Micro-Arcsecond
Scintillation-Induced Variability) survey of 443 flat spectrum sources
revealed that the IDV is related to the radio flux density and
redshift.
A study of the physical  properties of these sources has been severely
handicapped by the absence
of reliable redshift  measurements for many of these objects. This
paper presents 79 new redshifts and a critical evaluation  of 233
redshifts obtained from the literature. We classify  spectroscopic
identifications based on emission line properties, finding that 78\% of
the sources have broad emission lines and are mainly FSRQs. About 16\%
are weak lined objects, chiefly BL Lacs, and the remaining 6\% are
narrow line objects.
The gross properties (redshift, spectroscopic class) of the MASIV
sample are similar to those of other blazar surveys. 
However, the extreme compactness implied by ISS favors FSRQs and
BL Lacs in the MASIV sample as these are the most compact object
classes.
We confirm that the level of IDV depends on the
5\,GHz flux density for all optical spectral types. We find that BL
Lac objects tend to be more variable than broad line quasars. The level of
ISS decreases substantially above a redshift of about two. The
decrease is found to be generally consistent with ISS expected for
beamed emission from a jet that is 
limited to a fixed maximum brightness temperature in the
source rest frame.

\end{abstract}


\section{INTRODUCTION}\label{intro}

The discovery of centimeter wavelength, short  term  intra-day variability 
in some compact, flat-spectrum extragalactic radio sources 
\citep{Heeschen1984,Heeschen1987} was of immediate and profound astrophysical 
consequence. This phenomenon, which was soon dubbed Intraday Variability 
\citep[IDV;][] {Wagner1995}, implied brightness temperatures 
up to $10^{21}$ K in extreme 
cases \citep{Kedziora-Chudczer1997} if the 
variations were 
intrinsic. Such enormous brightness 
temperatures present a serious challenge to the current paradigm for 
the physics of extragalactic radio sources.

However, there is now overwhelming evidence that IDV results primarily 
from scintillation in the turbulent ionized interstellar medium of our 
Galaxy, 
commonly referred to as interstellar scintillation 
(ISS)\footnote{In this paper we call the phenomenon of
rapid variable flux density IDV and its cause, scintillation, ISS.}.  
This explanation was proposed by \citet{Heeschen1987b} 
and has now been confirmed from several further lines of observational evidence.
Time delays have been detected between the arrival times of the intensity 
fluctuations from 
ISS sources at two widely-spaced telescopes 
\citep{Jauncey2000, Dennett-Thorpe2002, Bignall2006}. 
Further, annual cycles have been detected in the timescale of ISS  
sources \citep[e.g.][]{JaunceyMac2001,Rickett2001,
Bignall2003,Dennett-Thorpe2003}, 
a periodic modulation that most plausibly results 
from the change in relative velocity of the Earth and the scattering 
medium through the course of a year. 
Subsequently, a strong correlation has been found between the level
of IDV and Galactic emission measure estimated from H$\alpha$
measurements \citep{Lovell2008}. This shows ISS to be the
predominant cause of radio IDV.
Though  some  questions remain 
\citep{Jauncey2001, Krichbaum2002}, 
interstellar scintillation is the only 
reasonable explanation of these observations
\citep{Jauncey2001}.

Though an extrinsic (as opposed to intrinsic) origin for IDV points to 
less extreme physical conditions for the sources that exhibit this 
phenomenon, scintillators are still among the most extreme and active 
radio AGN known. For a source to scintillate, its angular size must be 
comparable to that of the first Fresnel zone \citep{Narayan1992} which 
implies microarcsecond angular sizes for screen distances of tens to 
hundreds of parsecs. Such a high resolution cannot be achieved by any other 
existing technique, including space VLBI.
Thus, interstellar scintillation has the potential to probe
within a few light months of the central black hole \citep{Bignall2003}. 
Further, inferred brightness temperatures of a few scintillators are well in 
excess of $10^{14}$ K \citep{Macquart2000} which implies Doppler factors 
of several hundred or more \citep{Readhead1994} and is significantly higher 
than seen in VLBI surveys \citep[e.g.][]{Lister2009,Ojha2010}.
While Doppler factors very similar 
to those found in VLBI are implied from the ISS analysis of the Green Bank
radio monitoring program of 146 compact radio sources 
\citep{Rickett2006}, an investigation of the properties of AGN that exhibit 
ISS is of considerable astrophysical interest, both because 
their extreme physical 
properties make them ideal probes of existing models of AGN physics and they 
probe phenomena at unprecedented resolution.

The Micro-Arcsecond Scintillation-Induced Variability (MASIV) survey 
was undertaken by \citet{Lovell2003, Lovell2008} 
in order to construct a large sample of scintillating extragalactic sources.  
A key discovery from the MASIV survey 
is that both the number of sources showing ISS and their level 
of ISS variations
appear to decrease with increasing redshift \citep{Lovell2008}.
This result was obtained for the 275 (of 443) sources with measured redshifts.
While 206 of  these redshifts were obtained 
from the published literature,
69  of them were based on preliminary analysis of our unpublished 
optical observations.  A primary goal of the present paper is to 
publish the full analysis of those optical observations
and an additional 22 sources subsequently observed.

The apparent decrease in ISS with redshift is a new cosmologically important
result.   Its interpretation must include the cosmological
prediction that the angular size of an object should initially decrease
with redshift but start to increase above redshifts of about one.
Observational evidence for this phenomenon has long been sought, but
the best evidence to date is still only marginal 
\citep{Gurvits1999}.
The MASIV sources were selected for their flat radio 
spectra and many are quasars.
Thus any detailed interpretation must also address possible evolution in the 
beamed AGN jets presumably responsible for their very compact diameters.
Alternatively, there could be scattering effects in the inter-galactic medium
that could cause radiation from the more distant AGN to be scatter broadened 
so that their angular sizes are too great for them to scintillate in the 
interstellar medium of our Galaxy.

A follow-up to the MASIV Survey has
been conducted in which the ISS of a subsample of 140 MASIV sources
(70 with  $z<$ 2 and another 70 with  $z>$ 2) were
observed simultaneously at 5 GHz and 8.4 GHz over a duration of 11
days using the VLA \citep{Koay2011}. These observations 
provided a means of determining 
the origin of the redshift dependence of
ISS, by examining how the effect scales with frequency. This exploited
the fact that intrinsic source size effects and scatter broadening
both have different frequency dependences. The analyses and results
were presented in \citet{Koay2012}.

Our investigation of scintillators has been greatly handicapped 
by the lack of source redshifts. The MASIV survey found more
scintillators  among the weaker radio source sample,
which also showed more extreme scintillation behaviour 
than their higher flux density counterparts.
Since redshifts have been measured predominantly for 
the stronger radio sources, we particularly need to measure redshifts
for the weaker sources.  Redshifts are essential to determine 
physical properties  including linear sizes,
accurate brightness temperatures and luminosity distributions.
Optical luminosities are also an important probe of the physics of 
Compton scattering. For high Lorentz factor jets, energy losses should 
be dominated by Compton scattering of the diffuse radiation field 
of the cosmic microwave background, starlight and reprocessed emission 
from the nucleus \citep{Begelman1994}. This energy will escape mainly at 
X-ray wavelengths but some is expected in the optical band. Thus 
redshifts are needed to calculate the luminosities in order to 
understand the total energy budgets of these extreme sources. 

In this paper we present new redshifts obtained with the 2.56\,m 
Nordic Optical Telescope (NOT) on La Palma\footnote{Based on 
observations made with the Nordic Optical Telescope, operated on 
the island of La Palma jointly by Denmark, Finland, Iceland, 
Norway, and Sweden, in the Spanish Observatorio del Roque de 
los Muchachos of the Instituto de Astrofisica de Canarias.} 
and the 5\,m Hale Telescope at Mount Palomar.
We also present redshifts 
obtained from the literature whose reliability could be ascertained
and optical magnitudes collated from the surveys and literature.
Further, we look for relationships between the redshift, 
spectral classification,
luminosity and the observed level of ISS variation and compare the results
with a simple model for ISS.

Throughout this paper we adopt a cosmology with
 $\Omega_m=0.27,\Omega_{\Lambda}=0.73$ and $H_0= 70$ km
s$^{-1}$ Mpc$^{-1}$.
We use the convention  $S_{\nu} \propto \nu^{\alpha}$
for flux density at frequency $\nu$. 

\section{ THE MASIV SAMPLE}\label{sample}

The MASIV sample has its roots in two radio catalogues: CLASS (Cosmic
Lens All Sky Survey; Myers et al. 2003) and JVAS 
\citep[Jodrell Bank VLA Astrometric Survey;][] 
{Patnaik1992,Browne1998,Wilkinson1998}. Both of 
these were targeted surveys of flat-spectrum radio sources using the VLA at 
8.4\,GHz in the `A' configuration. JVAS is effectively a bright sub-sample 
of the 
complete JVAS/CLASS sample, which consists of more than 11,000 sources with 
flux density down to 30 mJy \citep{Myers2003}. We note that the CLASS flux 
density  limit is well below the 100 mJy flux density limit chosen for the 
MASIV sample (see below). 

For MASIV observations, we wanted to target compact flat-spectrum
sources that were unresolved at 8.4\,GHz with the VLA and located 
north of the equator. As a first step, all CLASS sources with modeled
source sizes $<$ 50 mas and all JVAS sources with $ >$ 95\% 
of their flux density in an
unresolved component were selected. These selection criteria were chosen
due to the way the data were presented in each catalogue: for JVAS, the
peak and total VLA flux densities are listed; for CLASS the modeled
component size is presented. Both criteria select those sources which are
essentially unresolved point sources for short (0.5 - 1.5 minute)
snapshot observations with the VLA in `A' configuration (this is the highest 
resolution configuration of the VLA). Sources where the catalogues indicated 
the presence of
any confusing sources -- indeed, any emission outside of the unresolved
target source -- in the VLA field, were dropped from the sample, as
selecting each MASIV target field to consist of a single, isolated point
source greatly simplifies the analysis of variability. 
To obtain a flat-spectrum sample with spectral index 
$\alpha \geq -0.3$ ($S_{\nu} \propto\nu^{\alpha}$),
these sources were cross-correlated with the NVSS catalogue 
\citep[NRAO VLA Sky Survey;][]{Condon1998},
although we note that the observed flux densities in
these catalogues were non-simultaneous.

The sample was next divided into strong and weak subsamples of about
300 sources each with the strong sample consisting of sources with 
$S_{8.4\,GHz} > 0.6$ Jy and the weak sample consisting of sources  
with $0.103 < S_{8.4\,GHz} < 0.13$ Jy. Finally,  the sources were
selected to have uniform RA -- $\delta$ distribution in order 
to have the best possible coverage for the VLA
observations \citep[for details see][] {Lovell2003}.

MASIV's 5\,GHz observations were carried out on this `core' sample of 578
sources. Of these, 102
sources were removed in the analysis stage due to the presence of
structure or confusion
while one was removed due to an error in the initial sample selection
process. This left a
sample of 475 point sources common to all four epochs of MASIV
observations. 
However, as described in \citet{Lovell2008}, 32 of these sources were
used as secondary calibrators in two or more epochs, which effectively
removes their low-level (instrumental or real) variations in those
epochs, relative to the other sources. These 32 sources were excluded
from the final MASIV sample, leaving 443 sources for full analysis.

As the average flux densities of many sources
from the originally defined radio ``strong'' and ``weak'' samples vary
with respect to the original catalogued flux densities, we used the
first epoch VLA data \citep{Lovell2003} to set the 
dividing flux density to 0.3 Jy at
5 GHz. In the rest of the paper we refer to the radio strong sample of
229 sources (including the 32 secondary calibration sources for the
optical analysis only, Section 4) and the radio weak
 sample of 246 sources.

\citet{Lovell2008}
report which of these sources exhibited Intra Day Variability (IDV)
 and in how many epochs of observation were classified as variable. 
A structure function (SF) analysis of the flux density variations
was also done, which provides a quantitative measure of the strength
of the variation.  The flux density was normalized by its mean and the
average SF was computed from the 4 epochs 
and tabulated  for each source as $D({\rm 2 days})$, the SF at a 
time lag of two days. 
This quantifies the variations and was shown to be 
correlated with the interstellar electron
column density, estimated from H$\alpha$ emission on 
that line of sight. The correlation confirms that 
the variations are indeed predominantly due to ISS.
There are thus two ways of quantifying the ISS of a source, either
as the number of epochs in which it was classified as variable
or the 4-epoch mean $D({\rm 2 days})$, where we note that 
no significance should be
attached to values smaller than about $4\times 10^{-4}$ 
\citep[for details, see][] {Lovell2008}.

\section{REVIEW OF THE LITERATURE}

The main goal of this work is to present the spectroscopic 
identifications and redshifts of the flat-spectrum, 
extragalactic radio sources that make
up the MASIV sample.

\subsection{Redshifts from the Literature}

We searched for redshifts and spectroscopic identifications 
of MASIV objects in the literature starting from data repositories like 
the NASA Extragalactic Database 
(NED\footnote{\url{http://nedwww.ipac.caltech.edu/}}). 
Spectroscopic identifications and  redshifts were found for the majority 
(171 of 229) of the  radio strong sample sources.  
However, this is not the case with  the weak sample (62 of 246).
Most literature redshifts and spectroscopic identifications are from
the 12th \citet{Veron2006} compilation of known AGN, hereafter 
VCV12. Identifications for 27 objects are from CGRaBS 
\citep[Candidate Gamma-Ray Blazar Survey;][]{Healey2008} and 
the rest from sources found using NED.

For those objects which had
reported redshifts, we tried to locate their spectra and/or
information about their emission lines
(line identifications, equivalent widths) 
from which the redshift estimates were obtained.
We classified the
sources into categories A, B or C, based on the reliability of their
redshift estimates. Sources with redshifts obtained from multiple
reported emission lines or with spectra clearly showing multiple
emission lines were placed in category A. If only a single line was
reported with wide wavelength coverage, or the spectrum provided was
of marginal quality, the source was placed in category B (see comments 
in Table~\ref{DataZMag}). Redshift
estimates of sources for which the spectra were not available, but
where the process by which the redshifts were obtained were clearly
described, were also placed in category B, as were three 
sources where imaging studies have detected the host galaxy but
show featureless spectra. For these three objects, we adopted the imaging 
redshift which is estimated  
assuming the host galaxy is a ``standard  candle'' 
\citep[for details see][] {Sbarufatti2005}.
Finally, objects in which
the redshifts were indicated to be photometric, or where no further
information could be found on how they were obtained, were placed in
category C. 
Objects
with (nearly) featureless spectra, with conflicting reported
redshifts, or with redshifts estimated based on an intervening
absorption system, were all placed in category C. 
Comments
on archival redshifts that either were rejected or judged less
reliable are in Appendix A.

\subsection{Optical identification}\label{opticalident}

In order to define the sample for optical
spectroscopy and select appropriate observing resources, accurate
optical identification was a prerequisite.

For  optical identification we used the CDS client 
database\footnote{\url{http://cds.u-strasbg.fr/}}
to access
 three large area surveys,  
the Sloan Digital Sky Survey (SDSS),
the Guide Star Catalog 2.3 (GSC)  and the 
USNO-B1 Catalog.
The  SDSS DR5 \citep{Adelman-McCarthy2007} is the primary catalogue 
for optical identification.
It is complete down to $r$=22.5 ($\lambda_{eff} \sim$6230\AA) with 
astrometric accuracy of about 
0$\farcs$1 \citep{Pier2003} and covers essentially the Galactic 
gap (RA $\sim$12$^{h}$
$\delta \sim$30$\degr$) and some smaller patches altogether about 
8000 square degrees.
As a secondary catalogue we used the GSC 2.3 \citep{Lasker2008}
which is compiled from scans of  the POSS-II plates and has F magnitudes 
($\lambda_{eff} \sim$6500\AA)  down to  $\sim$20.5. This has  poorer  
photometric (0.13-0.22 mag) 
and astrometric accuracy (0$\farcs$3-0$\farcs$4) than the SDSS. 
For 16 sources which were not found from the GSC2.3 we used  
USNO-B1 R2-magnitudes \citep{Monet2003},
which  has poorer photometric accuracy than the GSC2.3 \citep{Sesar2006}.
Of these sources  ten are  faint  ($R2>$19.0),
just visible from the POSS-I O-plates, and four are 
bright extended targets ($R2<$11.3).  
The ten faint objects might be missing from the GSC2.3 
due to variability, 
 while the four bright, extended
objects may be missing due to the design of the GSC2.3.
Finally, for eight faint sources  we found optical magnitudes 
from the literature.

\begin{figure}
\epsscale{1.0}
%\plotone{magsep.eps}
\plotone{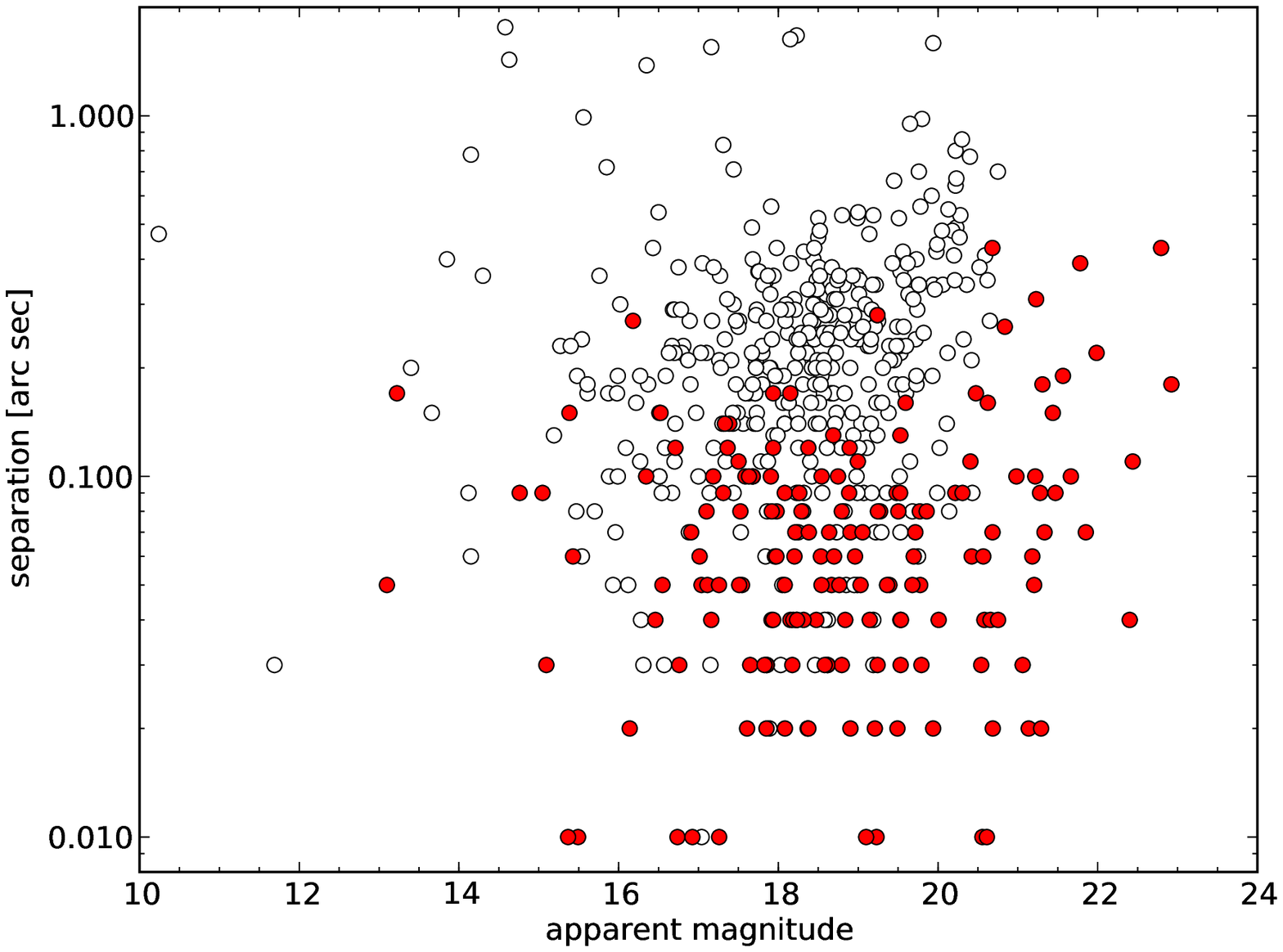}
\caption{Apparent SDSS $r$-magnitude (red dots) and GSC2.3 F-magnitude
(gray circles) versus separation  between the radio and optical positions.
\label{FigMagDist}}
\end{figure}

The initial search radius around the radio position was  5$\arcsec$.
From SDSS DR5 we found 165 matches 
to a MASIV radio source within one arcsecond, with 
 median $r$-magnitude of  18.7$\pm$1.8 (standard deviation of the 
 distribution) and the faintest optical identifications having $r$=22.9.
The mean separation increases from 0$\farcs$07 for 
$r<$ 20.5 mag sources to 0$\farcs$15 for the 37 sources with 
 20.5 $<$ $r$ $<$ 22.9. The GSC2.3 has  356 matches 
within 1$\farcs$5 with median magnitude of 18.4$\pm$1.5.
The mean separation of $F<$ 19.5 sources is 0$\farcs$25 and for the 
38 sources with $19.5<F<20.7$ it is 0$\farcs$39. 
Fewer than 10\% of the GSC2.3 matched sources have separations greater than 
0.5 arcsecond and only $\sim$2\% have separations greater than one 
second of arc.
In Figure~\ref{FigMagDist} 
we show a plot of the
measured separation versus apparent magnitude.
In both the SDSS DR5 and GSC2.3 matched sources, the accuracy of 
the astrometry is as expected and decreases close to the limiting 
magnitude of the survey.

Multiple matches were found in 14 cases ($\sim$4\% of the matches). 
Inspection of the data and 
visual inspection of the images (available from the CDS) revealed that close 
companions (8 sources), 
defined as those with a separation less than one arcsecond, 
were from separate epochs. This suggests that there is only a single 
source, but 
the astrometric solution is slightly different from epoch to epoch. 
For the remaining six sources the nearer object was selected as the 
optical counterpart and a second target was found with the 
angular distance 2.5-4.9 arcseconds from the radio source. 
Of the 356 GSC2.3 detected objects, 133 also have SDSS DR5 magnitudes, 
hence we are left with 223 sources with only F-magnitudes.
Down to a Galactic latitude of 5 degrees,
all  SDSS and GSC2.3 matches had only a single source 
within the search radius. 
Table~\ref{ObjOptId} summarizes the SDSS and GSC identifications.

In addition we defined a subsample,
($08^{h} < $RA$ < 16^{h}$ and $\delta < 64^{\circ}$), 
which covers roughly the SDSS DR5 sky area (hereafter `SDSS sample'). 
Of the 79 radio strong sources, all but one secondary calibrator source
have optical identifications. 
Optical identifications are available for 82 out of 95 sources in the
radio weak sample.

\subsubsection{Combining GSC2.3 and SDSS DR5 Data}

We transformed the  GSC2.3 and SDSS DR5 data to the 
common magnitude system using 
\begin{equation}
R_{SDSS} = r - 0.272(r - i) - 0.159
\end{equation}
by \citet{Chonis2008} for the SDSS data.
This is derived  using  SDSS DR5 and  \citet{Landolt1992}
data for $r>$ 14 magnitude stars. For  the  $r<$ 14 stars they 
found a systematic magnitude difference, 
however we have only two such bright
targets.

We obtained the GSC2.3 F- and N-magnitudes for 
the  standard stars, which were observed 
more than four times by  \citet{Landolt2009} where we
found (almost) a one to one correspondence
\begin{equation}
R_{GSC}=F.
\end{equation}
The  $F-N$ color term and zero point corrections are small
and are omitted (see Appendix B for details). 
The  USNO-B1 R2-magnitudes were converted in a similar way as 
the $F$ magnitudes. For 249 sources with 13 $<F<$ 19 the average 
magnitude difference $F-R2$ is $\sim$-0.07. 
Taking into account the
small number of $R2$-magnitudes and the magnitude offset,  
the possible bias to the global properties is negligible.

\begin{figure}
\epsscale{1.0}
%\plotone{magsep.eps}
%\plotone{Rsdss-RGSC.eps}
\plotone{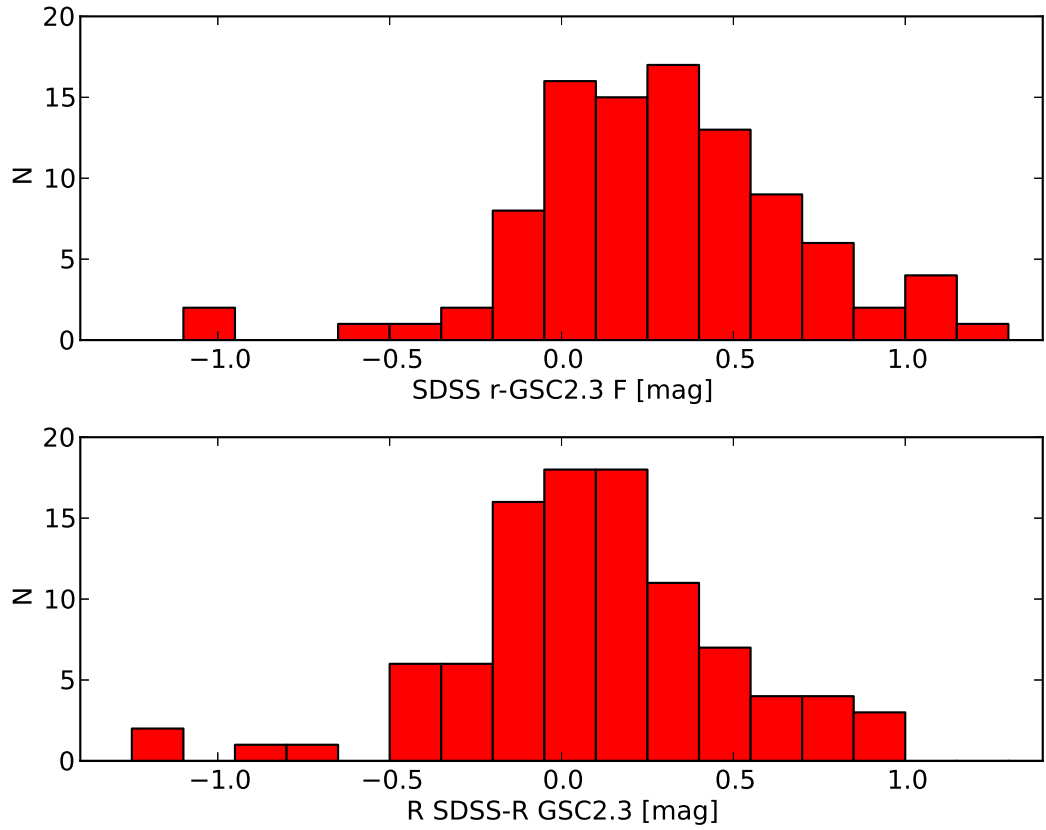}
\caption{The histogram of the magnitude difference for objects
with both SDSS and GSC identification.
The bottom panel shows the converted R-magnitude difference. 
\label{FigMagDiff}}
\end{figure}

We checked  the common magnitude system by selecting
all targets with SDSS photometry and 
GSC2.3 $F$ magnitude between 16.5 and 19.5, altogether
97 sources.  The median $r$ - $F$ magnitude difference is 
0.3$\pm$0.4 and after the $R_{SDSS} - R_{GSC}$  transformations
the median 
difference is 0.1$\pm$0.4 (Figure~\ref{FigMagDiff})
 suggesting that the systematic difference
between the transformed $R_{SDSS}$ and $R_{GSC}$ magnitudes
is small in comparison to the uncontrollables, such as 
variability of the sources and photographic plate sensitivity changes.

The magnitudes are integrated magnitudes and there 
has been no attempt to separate the host galaxy contribution from the 
nucleus or correct the emission line fluxes. 
For optically extended sources (all low redshift, $z\la0.2$
sources as well as  Sy2/NLRG-type objects  at higher redshift)  
the absolute magnitude
includes both the core and the host galaxy.
For example, J0057+3021
(NGC 315)  has USNO R magnitude $\sim$11.1 however
 the core  R-magnitude is $\sim$19.8 
 \citep[assuming V--R=0.3;][]{VerdoesKleijn2002}.
Also,  the GSC2.3 photometry pipeline was tuned for
point sources, thus there is a systematic overestimate for 
bright non-stellar magnitudes \citep{Lasker2008}.

All the magnitudes were de-reddened using the  $E(B-V)$ values and the 
relative extinction values from \citet{Schlegel1998}.
In  Table~\ref{DataZMag}  and for the optical luminosities, 
we refer to $R_{SDSS}$ and $R_{GSC}$ as $R$-magnitude.

\subsubsection{Radio  and  Optical  Properties}

The optical identification rate, combining the results from the 
SDSS DR5 and GSC2.3 catalogues, is $\sim$ 86\% 
for the radio strong and $\sim$ 77\% for the  radio weak
sample. However, since the two catalogues have 
different sky coverage and hence different Galactic latitude
coverage, this could bias
the identification rate.
Therefore, using only  the (all sky) GSC2.3  catalogue, 
the strong sample  has 
$\sim$ 80\% (187) of F-magnitudes brighter than 20 compared to  
$\sim$ 60\% (146) for the  weak sample.

Including empty fields\footnote{In the optical there is a well defined
upper detection limit, i.e. the median is a non-biased estimator.}, 
the median optical flux density
of the strong  radio sample is $\sim$0.15 mJy (18.2 mag) 
decreasing by a factor of 3.2 to $\sim$0.047 mJy (19.5 mag) 
for the weak sample. 
In comparison, the median radio flux density decreases by a factor of 
7.4, from $\sim$0.89 Jy to $\sim$0.12 Jy in the two samples. 
If we were to use this radio flux ratio 
to calculate the weak radio sample median optical magnitude from the 
strong radio sample mean optical magnitude, it would yield a
$R$-magnitude of 20.1 for the weak sample. 
The $k$-correction difference between the weak and strong 
samples is unknown, but most likely it is $\sim$0.2 magnitudes or less 
on average \citep[e.g.][]{Wisotzki2000}, assuming the 
average redshifts of 1.2 and 1.8 for the strong and  weak 
samples respectively (see below).

Figure~\ref{RadOpt} shows the optical versus radio flux density with 
diagonal lines indicating constant  radio-optical spectral index
(defined by $S_{\nu} \propto \nu^{\alpha_{ro}}$), which
has only been corrected for Galactic extinction and lacks any $k$-correction   
because 40\% of the radio weak sample objects have 
unknown redshift.
The top line corresponds to $\alpha_{ro}$=$-$0.2, the generally accepted 
radio quietness limit \citep{Stocke1985}. 
Although there appears to be no detailed correlation between optical
and radio flux density, the weak sample is on average one optical
magnitude fainter than the strong sample.  The radio weak sample
also appears to have an even larger spread than 
does the radio  strong sample.   The weak sample has 17 sources with 
$\alpha_{ro}> -0.45$  and the strong one only 
five. Close to $\alpha_{ro}$=-1.0  both samples have many optically 
empty fields thus we cannot draw any conclusions about the underlying 
distribution. It is also important to note that the non-simultaneous
nature of these radio and optical observations will further increase 
the scatter.
The difference
between the radio to optical flux ratios of the radio strong and weak
samples is statistically significant. Running the ASURV Kaplan-Meier
estimator \citep{Avni1980}, the probability that the two samples are
drawn from the same parent population is small ($P <$0.01\%).
This holds for the  `SDSS sample'  and also for the
whole sample using a shallower GSC2.3 limiting magnitude.

Our results indicate that the radio faint objects 
are optically ``brighter'' than
expected using the radio strong objects as reference. 
The explanation for the difference is not clear. One possibility is that the 
radio weak sources have larger host galaxy contributions in optical 
than radio strong objects.
Another possibility is that 
the average spectral energy distributions
of the strong and weak sources are different, perhaps
due to differing contributions of the 
accretion disk (``big blue bump'') to the optical flux density
or different amounts of dust.

\section{SPECTROSCOPIC FOLLOW-UP}

The new redshifts reported in this paper were obtained using the
2.56\,m Nordic Optical Telescope (NOT\footnote{
\url{http://www.not.iac.es}}) which is located at Roque
de los Muchachos, La Palma, Canary Islands, Spain and the 5\,m Hale
Telescope at Mount Palomar, California, USA. Most of the NOT data were
obtained during two observing runs (July 3rd to 7th, 2005 and July
22nd to 28th, 2006) supplemented by a few additional nights between 
January 2004 and July 2007. The Mount Palomar data were obtained during 
an observing run from August 9th to 10th, 2007. 
Most fields were pre-imaged at NOT with deeper and better image
quality than the SDSS frames to verify the optical identification of the 
radio source. Altogether  90 objects were observed at NOT and 
28 objects were observed at Palomar with some targets observed 
at both telescopes. 

At NOT, low resolution spectra were obtained using the Andalucia Faint
Object Spectrograph and Camera (ALFOSC\footnote{
The data presented here have been taken using ALFOSC, which is 
owned by the Instituto de Astrofisica de Andalucia (IAA) and operated 
at the Nordic Optical Telescope under agreement between IAA and 
the NBIfAFG of the Astronomical Observatory of Copenhagen.}) 
with Grism 4 (300 rulings/mm,
3200--9100 \AA, dispersion of 3\AA/pixel) with second order blocking
filter giving wavelength coverage about 3800--9100 \AA.  However, in
practice the NOT red limit of the spectrum is reduced to about 
8000 \AA~\ due to strong fringing.  We used Grism 6
(600 rulings/mm, 3200--5550 \AA) 
 or Grism 3  (400 rulings/mm, 3200--6700 \AA) 
for objects with a single detected
line at about 5000 \AA~\  in order to confirm the redshifts from 
either MgII, CIII] or the Ly-alpha line.  The NOT
run in 2005 was photometric, based on Carlsberg Meridian Telescope
(CMT) 
data\footnote{\url{http://www.ast.cam.ac.uk/$\sim$dwe/SRF/camc$\_$extinction.html}},
with seeing of 0$\farcs$6-1$\farcs$4 and the 2006 run had some dusty
and/or non-photometric nights with 0$\farcs$7-1$\farcs$1  seeing.  At
Hale, the Double Spectrograph (DBSP) was used with a setup of 600
lines on the blue arm (3400--5400\AA, $\sim$1\AA ~\ per pixel) and 158
lines for the red (5400--9500\AA, $\sim$4.9\AA ~\ per pixel) with
either 1$\farcs$0 or 1$\farcs$5 slit giving continuous wavelength coverage
from about 3400 to 9500 \AA.

Most of the targets were observed with the slit in parallactic angle
in order to reduce the light losses in blue.  
We note that, each spectroscopic target has a characteristic AGN spectrum 
and there is no evidence of any spurious counterparts.
In two cases the nearby companion was put on the slit and in both cases  
the object closer to the radio source 
had a characteristic AGN spectrum while the 
companion had a typical stellar spectrum. 
For the NOT data, an internal flat field (Halogen) image
was obtained before each science exposure,
in order
to improve the fringe correction at the red end.  In addition, several
HeNe images were taken during the night for the wavelength
calibration.  The DBSP calibration frames, dome flats, HeNeAr (red
arm), FeAr (blue arm) images, were taken morning and evening.  The
data were reduced using standard IRAF\footnote{IRAF (Image Reduction
and Analysis Facility) is distributed by the National Optical
Astronomy Observatory, which is operated by the Association of
Universities for Research in Astronomy (AURA) under cooperative
agreement with the National Science Foundation (NSF).}  procedures
including bias subtraction, flat field correction using internal flat
field lamp and wavelength calibration. Before extracting the spectrum, 
cosmic rays were removed from individual spectrograms.
The majority of the  spectroscopy
targets have R-magnitude between 17 and 21, with a median of
18.8.

The redshifts were determined from the narrow emission lines
(e.g. [\ion{O}{3}]$\lambda \lambda$ 4959, 5007, 
[\ion{O}{2}]$\lambda$ 3727)
whenever possible,
or from strong broad emission lines otherwise, typically
\ion{H}{0}$\beta$, 
\ion{MgII}{0},  \ion{C}{3}], \ion{C}{4} or Ly$\alpha$.
The new redshifts are based on two or more line identifications.  
However, in four cases with wide spectral coverage and 
high S/N, we assumed the line to be \ion{MgII}{0}, as 
if the line were to be \ion{C}{3}],
we would have 
expected to detect \ion{C}{4}] or \ion{MgII}{0}.
In Table~\ref{DataZMag} we list, respectively, 
source name, RA, declination, first epoch 5\,GHz flux density,
epochs of variability (at 5\,GHz), 
$D({\rm 2 days})$,
Galactic reddening,  de-reddened R-magnitude, 
 R-magnitude reference, redshift, spectroscopic identification
with the original  references and a comment.
Selected objects are discussed in Appendix A and C. Objects with 
only a single detected emission line are listed in
Table~\ref{ObjOneLine}.

\subsection{Classification Scheme}

On the basis of the optical spectrum the sources have been divided into 
three main groups: Type 0 objects with weak  emission lines
(mainly BL Lacs), Type 1 objects with  strong and broad  emission lines 
(mainly FSRQ) and Type 2 objects with strong and narrow  emission lines 
(mainly Seyfert 2's). In the context of AGN unification schemes, Type 0 and 
Type 1 objects have their Doppler boosted
relativistic jet aligned close to the line of sight
with Type 2 objects 
seen at a larger angle in such a way 
that the broad line region is obscured by a dusty torus \citep[see 
review by][]{Urry1995}.
The spectroscopic classification scheme is adapted from previous works
e.g. \citet{Laurent-Muehleisen1998}, \citet{Caccianiga2002b} 
and \citet{Caccianiga2008}.
Many observed properties of the AGN, 
such as emission line equivalent widths ($EW$) and luminosities
have continuous distributions. Hence the division between, for instance, 
Type 0 and Type 1 based on equivalent widths
$EW$ or between Seyfert 1 and FSRQ based on
optical luminosity depend on the dividing line chosen.
The VCV12 classifies all objects fainter than
$M_B=-23$ (using  VCV12 cosmology and magnitudes) as (low luminosity) AGN,
which is converted to our cosmology and 
R-band resulting in a QSO vs radiogalaxy (RG)
limit $M_R=-22.8$. 
As noted by the VCV12, the   $M_B =$--23
is an arbitrary limit without any physical reason. 
Due to different source classification schemes,  and/or to
variability,  a QSO might have 
been classified as an AGN in one epoch  but as an FSRQ 
at another epoch, and vice versa.
For example, a strong radio source, J\,1506$+$3730 is classified as an AGN 
by \citet{Healey2008},
NLRG by \citet{Sowards-Emmerd2004} and Sy2 by VCV12. 
Also, in some cases the classification depends on S/N, e.g.
a weak AGN-core with dilution by the host galaxy where broad emission 
lines might be undetected; 
for J\,0057$+$3021, \citet{Ho1995} reported a broad $H_{\alpha}$ line, but
a later study could not confirm it \citep{Goncalves2004}.

We describe our classification scheme below. Note
that all the line equivalent widths (EW) are rest frame values:\\

\noindent
0a) BL Lacs: objects with no lines or line rest frame 
$EW < $  5\AA\, and $C<25\%$,
where  contrast C is defined as 
\begin{equation}
C=(f_+ - f_-)/f_+,
\end{equation}
 where $f_-$ is  the mean flux
in the 3750--3950 \AA\, range and $f_+$ in the 4050--4250\AA\, range
 where flux is measured in frequency. 
In the case of moderate S/N we give 
only tentative spectroscopic identification.
\\
0b) BL Lac candidates: $25 < C <40 \%$  and strongest line $EW < 40$ \AA.\\
0c) Passive elliptical galaxies (PEG): $C>40\%$ and  $EW < 40$\AA.\\
1a) Flat spectrum radio quasar (FSRQ): at least one permitted 
line excluding $H\alpha$, line FWHM $>$ 1000 km/s and $M_R <$ -22.8.\\
1b) Seyfert type 1 galaxies: $M_R >$ -22.8, the subclass has been defined based 
on the  $H\beta$/[\ion{O}{3}]$\lambda$5007  line ratio \citep{Winkler1992}.\\
1c) Narrowline Seyfert1/QSO (NLSy1/NLQSO): 
At low redshift:   $H\beta$/[\ion{O}{3}]$\lambda$5007 $<$ 3 and
$H\beta$  $<$ 2000 km/s.
For the $z>1.5$ targets: rest frame line width 1000-2000\,km/s and 
the relative
strength of the \ion{He}{2} emission line compared to 
 \ion{C}{4} following  \citet{Heckman1995, Caccianiga2008}.
Type 1 objects have weak \ion{He}{2} emission line.\\
2a) Narrow line Radio galaxy (NLRG): line FWHM $<$ 1000 km/s
and  $M_R <$ -22.8.\\
2b) Seyfert type 2 galaxies:  $M_R >$ -22.8, the subclass has 
been defined based 
on the  $H\beta$/[\ion{O}{3}]$\lambda$5007  line ratio
\citep{Winkler1992}.\\
2c) Galaxy:  $C>40\%$, absorption lines e.g. \ion{Mg}{1} and no emission lines

Low redshift sources are the most problematic to classify as host galaxy 
dilution 
can hide broad lines and can reduce the line EW. Also the non-thermal 
continuum can 
be difficult to detect. Fortunately only four
of our 
follow-up targets are at low redshift ($z<$ 0.2), 
suggesting that the bias from the uncertain spectroscopic 
identification is small.

\subsection{Summary of the spectroscopic follow-up}

We report 79 new redshifts with spectroscopic identifications,
7 new BL Lacs with featureless spectra, and twelve objects with a single 
emission line of which six have a tentative redshift
(Table \ref{ObjOneLine} and Appendix C). We
confirm the BL\,Lac status of nine objects, and confirm the archival 
redshifts as well as spectroscopic identifications for nine other objects.
We also  re-classify two BL\,Lac objects using our own data
(Appendix A).  
Spectroscopic identifications from the literature were reviewed 
using the same scheme as for our own data, whenever possible.
In some cases this led us to adopt a different 
classification than NED (see appendix A).
Example spectra of our own observations are shown in Figure~\ref{ExamSpec}.

\subsection{RESULTS AND SAMPLE PROPERTIES}

Here we present 347 MASIV sources which have 
their spectra classified, either from the literature or based on our 
own data (Table \ref{DataZMag}).
The radio strong sample is 91\% (208/229) complete and the radio weak
sample is 57\% complete (139/246). Of the twenty-one 
radio strong sources without 
spectroscopic identification, seven have magnitudes from the 
surveys or literature, with R-magnitude between 17.3 and 23.99.

In Table~\ref{SpecIdSumm} we summarize the number of sources 
in each optical spectral classification, separating the radio strong 
and weak subsamples. It shows that Type 1 sources are the most common,
comprising 78\% of  both the radio strong and weak samples.
About 18\% of the radio strong sources and 16\% of the radio weak 
sources have Type 0 spectra. A small minority of the sources
have Type 2 spectra, 4\% and 6\% of the radio strong and radio weak samples, 
respectively. It also  shows that almost all 
Type 1 sources are optically luminous
FSRQs. 
In Table~\ref{StypeTab} we compare the identifications 
based on the present sample and the `SDSS sample'. 
The results suggest that the two samples have similar
distributions of the sources for both 
radio strong and weak subsamples. Also the redshift distributions
are identical.
This suggests that the observed sample 
is representative in comparison to the full 475 object sample.

Though broadly similar, there is some difference in 
the spectroscopic identifications of the radio strong and weak samples.
The radio weak sample appears to have slightly more low luminosity AGN
(e.g. PEG, LINER, HII-region, Galaxy) 
and fewer BL Lac objects than the radio  strong sample.
An increasing number of low luminosity objects are expected to be detected 
as the flux limit of the sample decreases.
At low redshift ($z<0.2$) most (7 out of 8) of the radio strong sample objects 
are BL Lacs or have broad emission lines, in contrast to the weak
sample, where half (6 out of 12)  of the $z<0.2$ objects have 
narrow emission lines or ``galaxy'' type  spectra and a further five
are classified as PEG.  Also, the PEG and galaxy type 
objects are absent in the radio strong sample.

All together, redshift information is available for 319 sources.
Figure~\ref{zHist}
shows the redshift distributions of subsamples divided by 
the optical spectral classification and the MASIV 
first epoch 5\,GHz flux density.
For Type 1 sources the median redshift increases from 1.3 to 1.4 
as the median  (5\,GHz) flux density decreases from 0.86 to 0.12 Jy.
The Type 1 radio strong sample has a clear maximum in the distribution,
but the  radio weak sample has a fairly flat distribution between
0.7 and 2.0. 
Of the Type 1 sources $\sim$20\%  of the 
radio strong and 
$\sim$29\% for the radio weak sample have $z>$ 2.0.
The redshift trend due to the flux density limit 
is similar to that seen from the previous blazar surveys
 e.g. from S5 $z_{mean}$= 1.18 (limit 1\,Jy) to S4 $z_{mean}$= 1.3
(limit 0.5\,Jy) to DXRBS $z_{mean}$=1.56 (limit 0.05 Jy)
 \citep{Landt2001}.
Most of the Type 0 or Type 2 sources  have measured redshifts  of 
less than one
and almost 50\% of the Type 0 sources are without a redshift.

We note that there might be a small number of mis-classified objects.
As in every survey, the signal to noise ratio of the spectroscopic data is 
not always optimized
for spectroscopic classification but for measuring the redshift. 
However, in most cases
there is no ambiguity in our source classification.
As our main results and conclusions are based on Type 1 sources (see below),
the effect  of some  mis-classified objects
is not significant.

\subsection{MASIV and  other Blazar Surveys}

In Table~\ref{BllFsrqOther} we compare spectroscopic 
identifications, limiting magnitudes and 
the radio flux density limit of MASIV with some other blazar surveys.
The  DXRBS sample \citep{Padovani2007a}, is selected by cross-correlating
X-ray sources with radio sources ($S_{5GHz}\ga$ 50mJy).
The CGRaBS \citep{Healey2008} has flat spectrum radio sources selected from the 
CRATES \citep{Healey2007} survey with $S_{4.8 GHz}>$  65 mJy, 
where the aim is to 
provide a catalogue of likely $\gamma$-ray loud AGN.   
This selection is based on a figure of merit (FoM) 
which measures the likelihood that an individual radio/X-ray source
is associated with a known  $\gamma$-ray source.
In terms of selection criteria the CLASS blazar survey (CBS) 
\citep{Marcha2001, Caccianiga2002b} is most similar to  MASIV. 
The CBS  has flat radio spectrum sources with the 
weakest sources $S_{5GHz} \ga $ 30mJy and red magnitude $\le$ 17.5.
It is interesting to ask how the source identification differs
between the MASIV selection criteria in comparison to the 
previous blazar surveys.

The DXRBS classification 
into the categories of ``Radio Loud Quasar, BL Lac and Radio Galaxies'' 
is almost 
identical to our Type 1, Type 0 Type 2. The possible difference is that
they might 
classify ``PEG'' as ``Radio Galaxy'' and their survey is 
nearly complete
\citep[$\sim$95\%,][]{Padovani2007a}.
The CGRaBS classifies objects as FSRQ (Type 1), AGN (Type 1),
BLL(Type 0), NLRG(Type 2) and GALAXY (Type 2) where 
our corresponding classifications are in parenthesis.
CGRaBS is about 79\% complete
with respect to the entire survey and 85\% for objects with known R-mag$<$ 23. 
The comparison between the 
CBS \citep[$\sim$91\% complete;][]{Caccianiga2004} 
and MASIV is straightforward, 
as the MASIV classification is adopted from the CBS.

The distribution of the spectroscopic classes is almost identical between
MASIV and the DXRBS and the CGRaBS with slightly more broad line
quasars in MASIV than other surveys.  
The main difference between the CBS and MASIV is  that 
the CBS has more low redshift Type 2 objects than MASIV.
This difference remains when using similar radio (`CBS rb') 
and optical  flux density (`MASIV ob') limits for these samples. 
This indicates that the selection of compact radio structure in 
defining the MASIV sample, 
filters out nearby Type 2 objects and favors FSRQs. 
Probably as a result of its bright optical flux density limit, 
the CBS has at least 40\% of low redshift ($z<$0.15) sources. 
In contrast, MASIV has 4\% for the total sample, 1\% and 
6\% for the radio strong and weak samples, respectively.
In all these cases it is assumed that the BL\,Lacs without redshift have 
$z>$0.15. 
Another difference between MASIV and the CBS is that 
sources that are optically less luminous but powerful in the radio 
(Figure \ref{rPowoLum} top panel, where  $M_R>-23$ and $P>10^{26}$ W/Hz)
are missing from the optically bright CBS sample.

The source and the redshift distributions 
of the MASIV, DXRBS and CGRaBS samples are similar. 
There might be slightly fewer $z >$ 3 sources in CGRaBS ($\sim$ 3\%) than in
MASIV ($\sim$5\%). The difference could be, for example, due to
incompleteness of the MASIV sample or pre-selection of only compact
radio sources for the MASIV sample.
MASIV has fewer  low radio power FSRQ sources ($P_r <$ 10$^{25.5}$ W/Hz) 
than  DXRBS  
 \citep[$22\%$;][]{Landt2001},
but about the same amount ($\sim$2-3\%) as the combined 1-Jy and S4 samples. 
Some of the sources classified as  ``galaxy'' might have weak broad lines
which are not detected due to 
dilution by the host galaxy continuum and/or
insufficiently high signal-to-noise ratio of the data.

Apart from decreasing the number of low radio power sources, using 
compact radio core selection increases
the number of objects in the MASIV sample that have featureless 
optical spectrum i.e. are classified as BL Lacs.
Specifically, it is interesting to note that IDV4 (i.e. IDV detected in all four
MASIV epochs) is one of the 
most effective preselections to find BL Lac objects. The use of 
persistent IDV 
as a selection criterion strongly favours BL Lac objects with 
43\% of the MASIV sources showing IDV in all four epochs identified 
as BL Lac objects. 
In comparison,  a radio--X-ray selection in the XB-REX sample 
\citep{Caccianiga2002a} and DXRBS \citep{Padovani2007a}
have a BL\,Lac fraction of  25\% and  18\% respectively.
However, $\gamma$-ray 
loudness may be an even stronger selector of BL Lacs with about 50\%
of Fermi/LAT identifications being BL\,Lacs \citep{Abdo2009}. 

\citet{Lister2001} studied the IDV properties of 
the Pearson-Readhead compact extragalactic radio sources and found that IDV
sources have smaller emission line widths and lower 5\,GHz luminosites than
non-IDV sources.  
Their results about line widths are consistent with our findings
that 25 to 40\% (for IDV1 through IDV4 subsamples) of the MASIV IDV
sources  are BL Lac objects and that 70\%
of the BL Lacs are IDV sources.  Whether there is a
correlation between IDV and line width 
amongst the Type 1 sources will be studied in
a future paper.
The 5\,GHz luminosity difference found by Lister et al.  could be 
due to combining FSRQs and BL\,Lacs into one IDV/non-IDV sample, and 
the different redshift distributions of the IDV and non-IDV samples.

\section{Discussion}

\subsection{Dependence of ISS on 5 GHz flux density
and optical spectral type}

The correlation between IDV and the 5\,GHz flux density
was discovered in the MASIV survey \citep{Lovell2003} and later
confirmed by \citet{Lovell2008}. In the latter work, the variability
of the flux density was studied using two complementary methods
(see Section 2):
$D({\rm 2 days})$  which combines all the 
data in a statistical estimator;  and a
visual analysis of each light-curve at each epoch, which classified
each source at each epoch as variable or not variable and so allowed
an estimation of the duty cycle in the variability.  Sources were
classified as ``ISS'' if they were found to be variable in 2 or more
epochs and as ``non-ISS'' otherwise.

To study the flux density/ISS correlation in detail, 
we divided the sample
by the flux density and spectroscopic type.
Figure~\ref{D2dHist} shows the
$D({\rm 2 days})$ histograms of the MASIV subsamples. There is a clear increase
of the median and mean $D({\rm 2 days})$ as the 
flux density decreases and this is seen
in all spectral types separately.  One can see that many of the
distributions in Figure~\ref{D2dHist} are significantly skewed, having a tail
extending to high values.  
When binned on a linear scale the distributions are more strongly skewed.  
A measure of  this skewness
is that the mean values of $D({\rm 2 days})$ are considerably larger than the
median, since the rare high values influence the mean but have little
effect on the median.

The Kruskal-Wallis (KW)
test indicates a $\sim$0.4\%  probability that the radio strong and radio weak
Type 0 samples have equal median $D({\rm 2 days})$.
The probability that Type 1 radio strong and weak samples 
have similar median is
negligible 
($p_{KW}<0.01\%$).
Similarly, the Kolmogorov-Smirnov (KS) test indicates that the
Type 0 radio strong and weak distributions of $D({\rm 2 days})$ 
are drawn from the
same parent population with a probability of 0.1\%; for Type 1 sources this
becomes $p_{KS} <$ 0.01\%.
For Type 2 objects the KW and KS tests indicate the 
same median or parent population
with less than 5\%  probability, however the full sample contains 
only 19 objects and their $D({\rm 2 days})$ values are typically at or
below the threshold (0.0004) for significant variation.
This indicates that the $D({\rm 2 days})$ 
has a $S_{5GHz}$ dependence and this holds 
for Type 0 and Type 1 sources with high statistical confidence. 
A similar, but less significant, trend is also seen for Type 2 sources.
We note that we have similar results when comparing 
the 5\,GHz flux density distributions for sources drawn from the 
`SDSS sample'. 
Thus we confirm the IDV--$S_{5GHz}$ flux density dependence found earlier 
and show that this correlation is present at least for 
Type 0 and Type 1 sources.

In previous studies of IDV, the optical  spectroscopic type of the objects  
has not been considered.
Figure~\ref{D2dHist} and Table~\ref{D2dTable} suggest that the IDV properties 
depend not only on the 5\,GHz flux density, 
but also on the optical spectral type.
Although the statistical tests
indicate a clear difference between the distribution of $D({\rm 2 days})$ in
the Type 0 and Type 1 sources for the full sample, there are
interdependencies that could cause an apparent correlation.

To study the apparent correlation between spectral
type and ISS, we first ask if
the different redshift distributions of Type 0 
and Type 1 sources could cause the  difference in the $D({\rm 2 days})$.
From previous studies, BL Lacs and FSRQs have been found to
have different observed redshift distributions \citep[e.g.][]{Giommi2012} 
with BL Lacs having a lower median redshift than FSRQs.
As many (25 out of 59) of our Type 0 sources are without known redshifts, 
it is probable that the redshift distributions of the
Type 0 and Type 1 sources are different. Hence it is
 difficult to compare Type 0 and Type 1 sources accurately.

First we compared the $D({\rm 2 days})$ distributions of Type 0 sources
with known redshift ($z_{max} \sim$ 1.15, N=16) to Type 1 sources 
with $z<$1.15 (N=57).
The median $D({\rm 2 days})$ is higher (0.00127 vs 0.00041) for
the  radio strong Type 0 sources  than Type 1. 
The statistical tests indicate some difference between the samples;
$p_{KS}=0.4\%$ that the distributions are similar, and 
$p_{KW}=2.5\%$ that the medians are the same.
When the Type 1 redshift cutoff is increased, the difference between 
the two samples increases. 
No comparisons were made using sources only from the SDSS
sample, as there are only nine Type 0 sources; however the median
D(2days) of Type 0 sources is higher than that of Type 1 sources.

Next we compare all Type 0 radio strong sources 
(N=31, median $D({\rm 2 days})$=0.00070)
to radio strong Type 1 sources  with different redshift cutoffs. 
The KS test indicates $p=2.3\%$ probability that the Type 1 
with $z<$1.15,
and  Type 0 $D({\rm 2 days})$ distributions
 are drawn from the same parent population and the 
KW probability that the medians are the same 
is $p_{KW}=2.3\%$.
When the Type 1 redshift cutoff is increased, the KW and KS probabilities
that the medians or distributions  are similar decreases.
For example, comparing all Type 0 radio strong sources to radio strong
Type 1 sources with $z<$1.3 the $p_{KW}=1.1\%$ and $p_{KS}=0.5\%$.
Considering only
radio strong sources from the SDSS sub-sample, the median 
$D({\rm 2 days})$  of the
Type 0 is also higher than that of the Type 1 sources, but the difference
is not statistically significant for these smaller samples.

For the radio weak samples, including all Type 0 and Type 1 sources,
there is a statistically significant difference between the two samples
($p_{KS}=0.3\%$ $p_{KW}=1.8\%$). However, including only $z<$ 1.5
Type 1 sources the distributions are similar. Finally, excluding the 
low redshift ($z<$ 0.16) Type 0 sources, mainly PEG, and 
including only $z<$ 1.5 Type 1 sources the statistical tests
indicate significant difference between the two samples
($p_{KS}=0.3\%$ $p_{KW}=2.1\%$).

Using the current data, we conclude that the Type 0 sources 
have higher $D({\rm 2 days})$
values than Type 1 sources. However, the statistical significance
is dependent on the sample selection, especially the redshift cutoff 
of Type 1 sources. 
As many Type 0 sources are without known redshift, it is 
impossible to draw solid conclusions regarding whether 
there is a real $D({\rm 2 days})$ 
difference between the Type 0 and Type 1 sources.
In general, the  redshift distributions of
BLL (Type 0) and FSRQ (Type 1) are different
\citep[see e.g.][]{Giommi2012},
with BL Lacs having a lower median redshift
than FSRQs.  \cite{Giommi2012} proposed that BLL and FSRQ would have
similar intrinsic redshift distributions,
however, in this case the $D({\rm 2 days})$ distributions
of Type 0 and Type 1 do not match, suggesting different
$D({\rm 2 days})$ properties between Type 0 and Type 1 sources.

Table~\ref{VarEpochs} shows the number of sources of each spectral
type compared with the number of MASIV Survey epochs in which the
source varied according to the visual classification presented by
\citet{Lovell2008}. 
This suggests that Type 0 sources show ISS more often than the 
Type 1 sources, which is 
consistent with the $D({\rm 2 days})$ results above.
It is notable that Type 0 sources have the highest levels 
of ISS in both the weak 
and strong groups, which implies that they typically have more compact radio 
structure than Type 1 sources.
The ISS intermittency is likely to be source related, rather than 
purely due to ISM intermittency.
These results are relevant to attempts to understand the physical 
differences between the two groups.

\subsection{ISS versus Redshift}\label{SectISSz}

\citet{Lovell2008} found a redshift dependence in the ISS properties.
The original 275 radio-selected MASIV sources with redshifts revealed
that ISS of compact sources at 5\,GHz decreases with redshift 
above  $\sim$2. From the optical data presented here
we can examine whether this decrease depends
on spectral classification and re-examine the 
decrease with redshift using the larger sample
of 320 objects with redshifts.
As the  $D({\rm 2 days})$ correlates with the 5\,GHz flux
density, if the 5\,GHz flux density were to correlate with redshift, this
could cause an artificial  $D({\rm 2 days})$-redshift correlation.
We ran Pearson and Spearman correlation
coefficient tests for the full and `SDSS sample' 
radio strong and weak Type 0 and Type 1 sources. 
Out of eight correlation coefficients only one was
statistically significant, the full radio strong sample Pearson test
($r\sim$ -0.18; $p\sim$3\%). All the other tests indicated no correlation
between $S_{5GHz}$ and redshift.

The 2008 MASIV analysis
of $D({\rm 2 days})$  versus redshift included all sources regardless of their
spectral classification, specifically including Type 0 sources with known
low redshift and excluding the Type 0 sources with unknown, possibly
high, redshift.
We ask whether the presence of Type 0 sources
could be partly responsible for the drop in ISS with
increasing redshift. 
We found (see above) the Type 0 sources to be
among the strongest scintillators and also to be concentrated 
at redshifts less than 0.7, however  almost 50\% (27) of the 
sources do not have a redshift.
Fortunately, the
redshifts of the Type 0 sources are less than 1.15, so the possibly
misclassified Type 1 sources will most likely affect the low-redshift
end of the Type 1 sources.
\citet{Lovell2008} 
show that the average $D({\rm 2 days})$  values drop
around $z\sim 2$, hence the  misclassified Type 1 sources should have negligible 
bias to the ISS-redshift correlation.

We now restrict the analysis 
to the Type 1 objects which constitute the great majority of our sources.
Figure~\ref{D2dvsz} shows the $D({\rm 2 days})$ against redshift, with
horizontal lines indicating the median $D({\rm 2 days})$ 
for different redshifts bins.
Both radio strong and weak sources have the lowest $D({\rm 2 days})$ 
median at the highest redshift bin ($z>2.5$). Combining the radio strong 
and weak  samples, the KW test indicates 0.5\% 
probability that the $1.5<z<2.5$ and $z>2.5$
samples have similar $D({\rm 2 days})$ 
median and the KS-probability that the two samples are drawn from
the same parent population is 3.2\%. 
We also note that the two highest 
redshift bins ($D({\rm 2 days})$ = 0.00029 and 0.00051)
have about equal numbers of radio strong and radio weak sources, reducing
the probability that the $D({\rm 2 days})$
 decrease with redshift is induced by the
5\,GHz flux density.
Figure~\ref{D2dvsz} suggests that the radio weak sample has a
stronger redshift dependence than the radio strong sample.
The KW probability that the weak sample
$D({\rm 2 days})$ medians at $z>2.5$ and $1.5<z<2.5$ are the same 
 is 2.0\% and the KS for the same parent population is 1.4\%. 
From these results we conclude that Type 1 sources show a decrease in
$D({\rm 2 days})$ with increasing redshift, though the trend is not as strong
statistically as the decrease versus 5 GHz flux density.

\subsection{ISS and Radio Power}

We searched for correlations between radio power
and  the $D({\rm 2 days})$, and compared the radio powers
of ISS and non-ISS sources.
We  calculated the rest-frame radio power at 5\,GHz for
isotropic emission as 
\begin{equation}
P_r=4\pi D^2_L S_5(1+z)^{-(1+\alpha_{r})},
\label{RadioPower}
\end{equation}
where  $D_L$ is the luminosity distance in the adopted $\Lambda$CDM 
cosmology, $S_5$ is the mean first epoch MASIV VLA flux density and
the last term is for the $k$-correction.
We have assumed  $\alpha_{r}$=0  for all our sources.

As the initial sample selection has a
gap in the 5 GHz flux density distribution between about 0.2-0.5 Jy, the
radio power distribution is not continuous at a given redshift. In order to
study continuous distributions, we selected Type 1 sources with $1.0< z
<1.8$, 59 radio strong and 39 radio weak sources, and divided the sample
into ISS sources (with $D({\rm 2days}) > 4 \times 10^{-4}$) and non-ISS
sources ($D({\rm 2days}) < 4 \times 10^{-4}$) to compare the radio power
distributions. The KS and KW tests suggest significant difference in the
radio power distributions ($p_{KS} = 0.9\%, p_{KW} = 0.3\%$), non-ISS
sources being more powerful.
The Pearson correlation
coefficient indicates weak correlation ($r$=-0.28 $p$=0.5\%)
between radio power and $D({\rm 2 days})$.
However, as there are fewer radio weak  sources than radio strong
sources (Table~\ref{RadioLuminT}, Figure~\ref{rPowoLum}), 
this can cause a correlation between ISS and radio power. Thus
we divided the sample into radio strong and radio weak and 
compared the radio power distributions of the ISS and non-ISS sources. 
The KS and KW tests  show
no difference between the samples, suggesting that the 
correlation seen by the full sample is weak or non-existent.
The redshift distributions of the samples are identical in each case
and the results are not sensitive to the selected ISS/non-ISS limit. 
Our interpretation of the results is that the ISS-$P_r$
correlation is weak.

Finally, we note that using the radio loudness limit of $P_r$= 10$^{23.7}$ W/Hz,
\citep[similar to][]{Laurent-Muehleisen1998}
five low redshift ($z<$0.05)
sources are ``radio quiet''\footnote{J0248+0434, J1141+5953 and J1604+1744 
are classified as galaxies, J1112+3527 as PEG and J0057+3021 as a LINER}.

\subsection{ISS and  Optical Brightness}

In the optical, we searched for correlations between $D({\rm 2 days})$ and
optical magnitude/optical luminosity.
We selected subsamples, as above, based on the radio power
and carried out the Spearman and Pearson correlation tests.
Apart from Type 0 radio weak sources we found no evidence of 
correlation between the optical $R$-magnitude and $D({\rm 2 days})$. 
As several Type 0 sources have low
redshifts, and might have a strong host galaxy contribution to the optical
magnitude, we excluded the low redshift sources ($0.01 < z < 0.15$), and
the indication of correlation disappeared.
We calculated the rest-frame optical power as in equation~\ref{RadioPower}
but used the R-band flux density and 
a  $k$-correction $-2.5(1+$$\alpha_{o}$$)log$$_{10}$$(1+z)$, where 
we assumed  $\alpha_{o}$=$-0.5$ for all our sources.
The statistical tests (KS, KW, Spearman) indicate no 
difference in the optical luminosities of the ISS and non-ISS 
sources nor any correlation between the  $D({\rm 2 days})$ and 
optical luminosity.

\subsection{Modeling the ISS dependence on Redshift}

\label{ISS_z}

The 275 radio-selected MASIV
sources with redshifts known prior to the present paper revealed that 
ISS of compact sources at 5\,GHz decreases with redshift 
above $z \sim 2$ \citep{Lovell2008}. 
In Figure 3 of \citet{Koay2012} we considered the
possible influence of a cutoff in radio power or in redshift  and
found that either a single cutoff beyond redshift $z$=2 explains the
data,  or different cutoffs in $P_r$ would be required for the weak
and strong samples.   Here we follow the simpler hypothesis of a
physical model in which $D({\rm 2 days})$ is governed by $S_{5GHz}$ 
and redshift
regardless of $P_r$.  The model is motivated by the physics of jet
emission as discussed below and also by \citet{Koay2012}.
This model also explains why the radio weak sample has a
stronger redshift dependence of $D({\rm 2 days})$
 compared to the radio strong
sample \citep[see Figure 9 in][]{Koay2012}.

In Figure \ref{D_z} we have averaged $D({\rm 2 days})$ 
into four redshift bins for the FSRQs alone, combining both weak and strong
sources.  We note that excluding the Type 2 sources increases 
the level of ISS in the lowest redshift bin and excluding 
the Type 0 sources decreases 
the level of ISS in the lowest redshift bin, 
when compared 
with our earlier results \citep{Lovell2008}.  We also over plot the data with
curves from a simple theoretical model.   The model makes
many assumptions and so is quite simplistic; however, it has the 
merit of including the most obvious cosmologically important parameters.
We assume that the scattering is due to a single
layer of isotropic Kolmogorov turbulence in the interstellar electron density
centered at 500 pc from the Earth.   We further assume that the 
scintillation for a point source at 5\,GHz is at the transition 
from weak to strong scintillation.  
We model the compact part of the sources by a relativistic jet pointed 
toward the Earth emitting beamed radiation characterized by a brightness 
temperature in the jet frame of $T_{b0} = 10^{11}$ K which delivers 
an observable flux 
density of 1\,Jy.  This provides a crude model for synchrotron radiation that is
self-absorbed causing an upper limit in brightness.
The jet is characterized by a Doppler factor $\Gamma$ which is assumed
independent of $z$.
The apparent brightness temperature of the source at redshift $z$
is thus given by  $T_{b0} \Gamma /(1+z)$.

Since for a flux density limited sample the apparent angular diameter varies
as the inverse square root of the apparent brightness temperature, 
the (1+$z$)
reduction factor causes the apparent diameter of the compact source to
increase as $\sqrt{1+z}$,
which corresponds 
to the cosmological ``angular size versus redshift'' phenomenon.    
VLBI observers have long sought
to detect this predicted increase in angular diameter with redshift, but with
only partial success.  For example, \citet{Gurvits1999} show 
angular diameters of compact radio sources that exhibit a barely
significant increase beyond a redshift of about 1.  We see a highly 
significant drop in
$D({\rm 2 days})$ with redshift, but a model for the ISS phenomenon is 
required before
it can be interpreted cosmologically.  An ISS model was applied to 
the data from the 
Green Bank monitoring observations by \citet{Rickett2006}.
However, here we have improved the model by using the formulae developed by
\citet{Goodman2006}, which explicitly include contributions from 
diffractive scintillations,
that are important near the transition from weak to strong scintillation
and were not included by \citet{Rickett2006}.  An understanding of
$D({\rm 2 days})$ is obtained by noting its asymptotic dependence
$\propto \theta_{\rm source}^{-2}$ for intraday ISS.
This assumes that the ISS is quenched and that its time scale is 
not much longer than 2 days.  
However, if the time scales are much longer, then $D({\rm 2 days})$
drops even more steeply with $\theta_{\rm source}$.  The values chosen 
for $T_{b0}$, $\Gamma$ and the mean flux density $S_c$ in the compact 
source enter
through a single quantity $T_{b0} \Gamma/S_c$.  Thus, while the model exhibits 
the expected dependence on redshift, we should also expect a wide range in this 
unknown quantity.

A single model curve ($\Gamma = 20$) can be seen to overlap three of 
the four observed 
values within $\pm 1 \sigma$.  However, in the highest redshift bin,
$D({\rm 2 days})$ lies more than $2 \sigma$ below the curve.   
A steeper than predicted drop in ISS with redshift could imply redshift
 evolution in
the Doppler factor of the jets or angular broadening caused by propagation 
through the intergalactic medium, as mentioned by \citet{Lovell2008}. 
However, the discrepancy shown here is not statistically strong 
enough and we await
further redshift values from the remaining (weak sample) sources
before further analysis.  
In fact, this
discrepancy at the highest redshift bin could be linked to a
steepening of the mean source spectral index with increasing redshift,
coupled with the lower $D({\rm 2 days})$ of steeper spectrum 
sources, as has been
found in the dual-frequency follow-up observations \citep{Koay2012}.
As noted in Section 2, the flux densities used to estimate
the spectral indices for the selection of MASIV sources were
non-simultaneous; and the dual-frequency follow-up observations
provided more accurate estimates.
Indeed any more detailed analysis must 
also consider the
distributions in the parameters of the model rather than the
single values adopted here.  Nevertheless, we stress that the
present work presents the strongest observational evidence
to date for the predicted increase in angular size due to 
cosmological expansion.

\section{SUMMARY}

We present the optical and spectroscopic identifications 
of a sample of  475 flat-spectrum radio sources which are unresolved
at 5\,GHz (FWHM $\la$ 50 mas). About half of these sources also show
ISS. Our sample is divided into radio strong (S$_{5GHz}>$ 0.3 Jy) 
and radio weak ( 0.06$<$ S$_{5GHz}<$0.3 Jy)
categories. 
For optical identification the SDSS DR5 and GSC2.3 surveys were used. 
About 80\% of the radio strong and  60\% of the radio weak  
sources  have an optical counterpart with  $R<$ 20 magnitude.
Using sub-samples
defined from the SDSS data  indicates that the nearly complete radio
strong sample ($\langle S_{\rm 5GHz} \rangle = 0.9$\,Jy, N=78) has median
$r$-magnitude $\sim 17.9  \,(r_{\rm max} \sim 22.0$), and the 70\%-identified
radio weak sample has $r \sim$ 18.9  ($r_{\rm max} \sim$ 22.9, $\langle S_{\rm
5GHz} \rangle = 0.1$\,Jy, N=95).

Spectroscopic identifications and redshifts are from the literature 
and from our own observations (347 sources).  
Spectroscopic identification of the radio strong sample is 91\% 
complete and the weak sample is 57\% complete.

The key findings can be summarized as follows:

1) The radio weak sources have flatter radio-optical spectral index than
the radio strong sources. 

2) Almost 80\% of the MASIV sources are identified as broad line objects, 
$\sim$13\% as BL Lacs and $\sim$7\% as narrow line objects   
or galaxies. The spectroscopic identifications are  
similar for the radio strong and weak samples.  
Our results suggest that selecting for compact radio structure  
favors FSRQs and BL Lacs over other flat spectrum sources.

3) Of the scintillating sources, 25\% are BL Lacs and the  rest are almost 
exclusively broad line objects. The fraction of BL Lacs increases with 
the degree of persistence of ISS and 40\% of the 
sources with most persistent ISS
(scintillating at all four epochs) are BL Lac objects. 
This is
consistent with BL Lacs being more strongly core dominated than FSRQs, and
could possibly indicate relativistic jets that are pointed closer to the
direction of the Earth.
We confirm that $D({\rm 2 days})$ depends on 5\,GHz flux density, 
with weak sources having higher $D({\rm 2 days})$ values.
A similar trend is found also when broad line or weak line 
objects are studied separately.
We found indications that the ISS properties correlate
with emission line equivalent width.

4) Radio and optical luminosities of the
scintillating and non-scintillating broad line
sources are similar at a given redshift, suggesting intrinsically 
similar SEDs. 

5) We find no correlation between the optical luminosity and ISS.
We find a weak
correlation between radio power and ISS. 
However at given redshift 
the ISS and non-ISS sources have similar radio power.

6) For broad line objects, we confirm the sharp decrease in the number of ISS 
sources and in the level of their ISS above a redshift $\sim$ 2.
The decrease is compared to a 
simple model for ISS of flat spectrum radio sources with maximum 
brightness temperatures that are Doppler-boosted in jets 
pointing toward the Earth.
There is reasonable agreement but more redshifts are needed for the weak radio 
sample.
We found strong  observational evidence
for the predicted increase in angular size due to 
cosmological expansion.

\acknowledgments
We thank the referee for perceptive and constructive comments that have improved the paper. 
RO acknowledges the Access to Major Research Facilities Program (AMRFP) of the 
Australian Government for travel support (Grant Number: 05/06-0-04) for
 observations 
with the Nordic Optical  Telescope (NOT) during July 2005. 
We wish to acknowledge travel support
from the Access to Major Research Facilities Program which is
administered by the Australian Nuclear Science and Technology
Organisation. DLJ would like to thank the Research School of Astronomy
and Astrophysics of the Australian National University, for their
hospitality.  
BJR thanks the US
NSF for partial support under grant AST 05-07713 and for the
hospitality of the Cavendish Astrophysics group at Cambridge
University.
JYK is supported by the Curtin Strategic
International Research Scholarship provided by Curtin University.
This research was supported by an appointment to the NASA
Postdoctoral Program at the Goddard Space Flight Center, administered
by Oak Ridge Associated Universities through a contract with NASA.

This research
 has made 
use of the SIMBAD database, operated at CDS, Strasbourg, France. This 
research has 
made extensive use of the NASA/IPAC Extragalactic Database (NED) which 
is operated 
by the Jet Propulsion Laboratory, California Institute of Technology, 
under contract with 
the National Aeronautics and Space Administration. This research has 
made use of 
NASA's Astrophysics Data System Bibliographic Services.

Funding for the SDSS and SDSS-II has been provided by the Alfred
P. Sloan Foundation, the Participating Institutions, the National
Science Foundation, the U.S. Department of Energy, the National
Aeronautics and Space Administration, the Japanese Monbukagakusho, the
Max Planck Society, and the Higher Education Funding Council for
England. The SDSS Web Site is http://www.sdss.org/.

The SDSS is managed by the Astrophysical Research Consortium for the
Participating Institutions. The Participating Institutions are the
American Museum of Natural History, Astrophysical Institute Potsdam,
University of Basel, University of Cambridge, Case Western Reserve
University, University of Chicago, Drexel University, Fermilab, the
Institute for Advanced Study, the Japan Participation Group, Johns
Hopkins University, the Joint Institute for Nuclear Astrophysics, the
Kavli Institute for Particle Astrophysics and Cosmology, the Korean
Scientist Group, the Chinese Academy of Sciences (LAMOST), Los Alamos
National Laboratory, the Max-Planck-Institute for Astronomy (MPIA),
the Max-Planck-Institute for Astrophysics (MPA), New Mexico State
University, Ohio State University, University of Pittsburgh,
University of Portsmouth, Princeton University, the United States
Naval Observatory, and the University of Washington.

\clearpage

{\it Facilities:} \facility{NOT()}, \facility{Hale ()}.

\bibliographystyle{jwaabib}
\bibliography{mnemonic,aa_abbrv,masiv}
%\bibliography{mnemonic,jwaabib,masiv}

\clearpage

%\clearpage
\begin{figure}
\epsscale{1.0}
\plotone{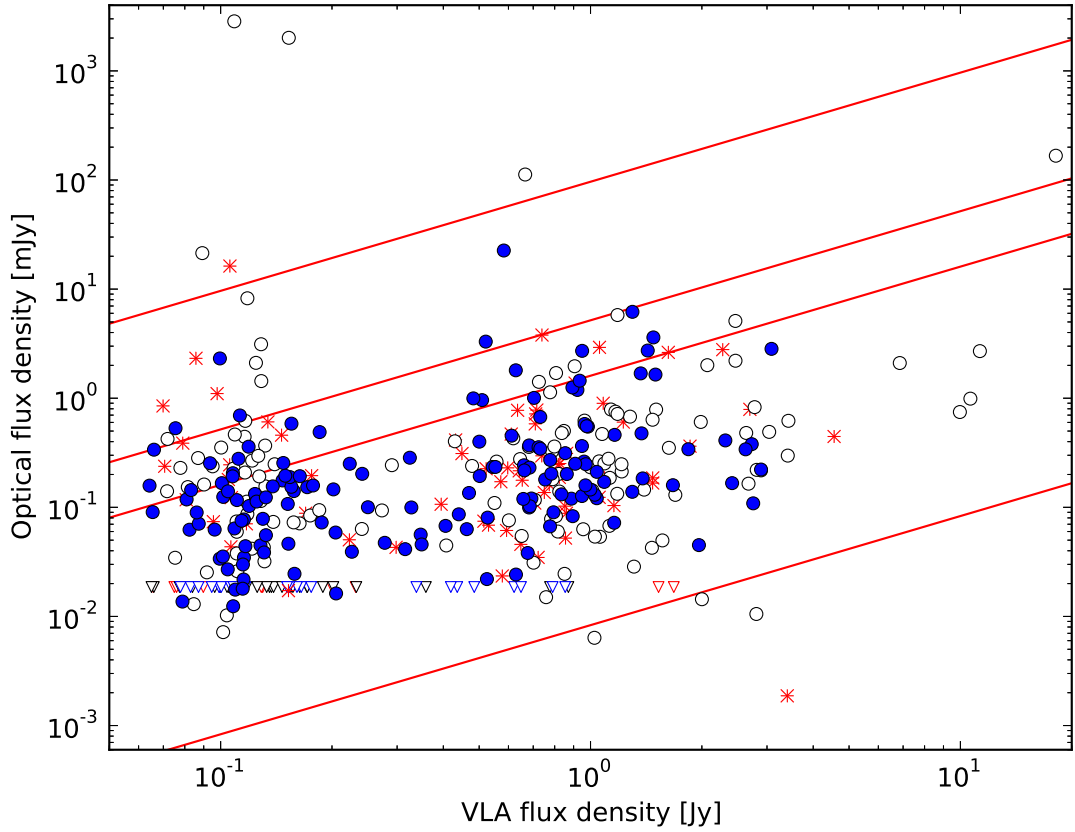}
%\plotone{vla_mag.eps}
\caption{
5\,GHz flux density vs optical flux. The blue circles, red stars
and open circles are for
variable, non-variable and  intermittently variable, respectively.
The magnitude upper limits (R=20.5) are indicated with blue,
red, black triangles for variable, 
non-variable and  intermittently variable, respectively.
The diagonal lines 
indicate constant radio--optical spectral index -0.2, -0.45, -0.55 
and -1.0  from top to bottom.
\label{RadOpt}
}
\end{figure}

\begin{figure*}
\epsscale{1.}
%\plotone{spec.eps}
\plotone{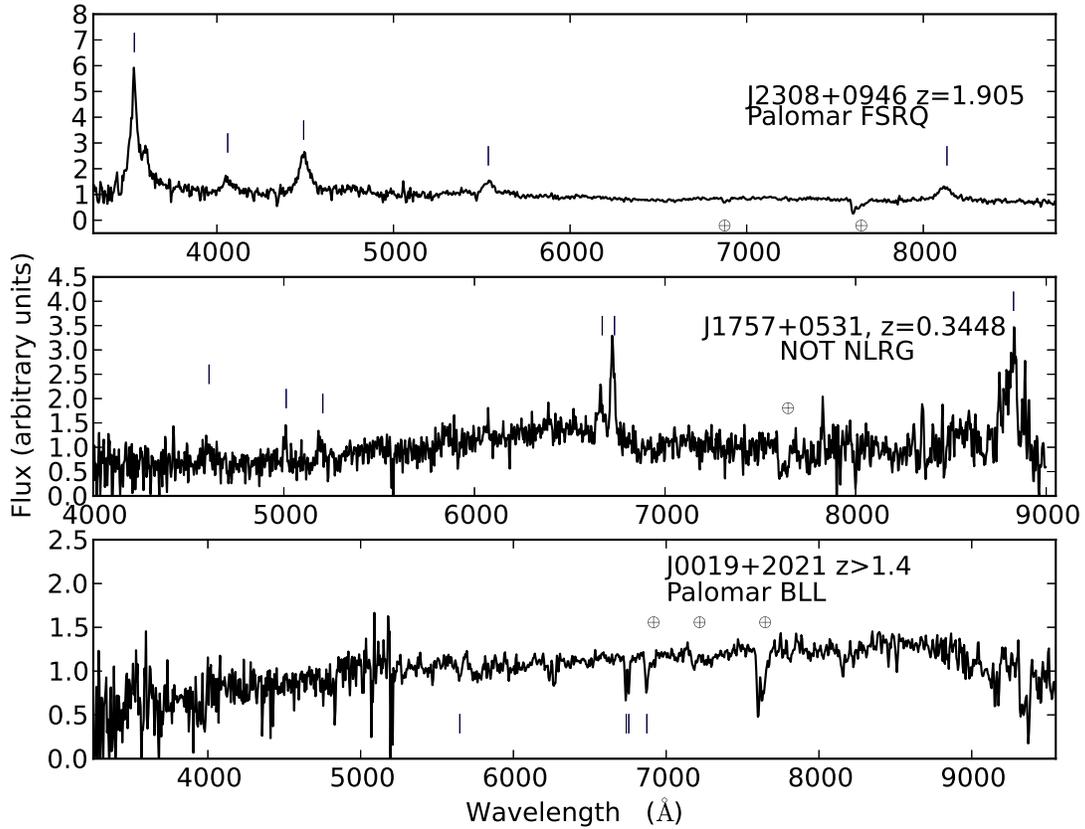}
\caption{
Example spectra of an FSRQ (Type 1), NLRG (Type 2) and  BL Lac object.
In the top panel the vertical bars indicate redshifted  
Ly$\alpha$, \ion{Si}{2},
 \ion{C}{4}$\lambda$1549, \ion{C}{3}$\lambda$1909 and 
\ion{Mg}{2}$\lambda$2798. In the middle panel vertical bars indicate
redshifted 
\ion{Ne}{5}$\lambda$3426, 
\ion{O}{2}$\lambda$3730, 
%\ion{Ne}{3}$\lambda$3870, 
\ion{Ne}{3}$\lambda$3870, 
[\ion{O}{3}]$\lambda \lambda$4958, 5007 and $H\alpha$.
In the bottom  panel the  vertical bars indicate 
absorption lines at redshift 1.41  with \ion{Mg}{1}$\lambda$2852, 
\ion{Mg}{2}$\lambda$$\lambda$2803,2796 and \ion{Fe}{2}$\lambda$2344.
The  $\oplus$-signs  identify features due to the Earth's atmosphere
\label{ExamSpec}}
\end{figure*}

\begin{figure}
\epsscale{1.0}
%\plotone{figz_hist.eps}
\plotone{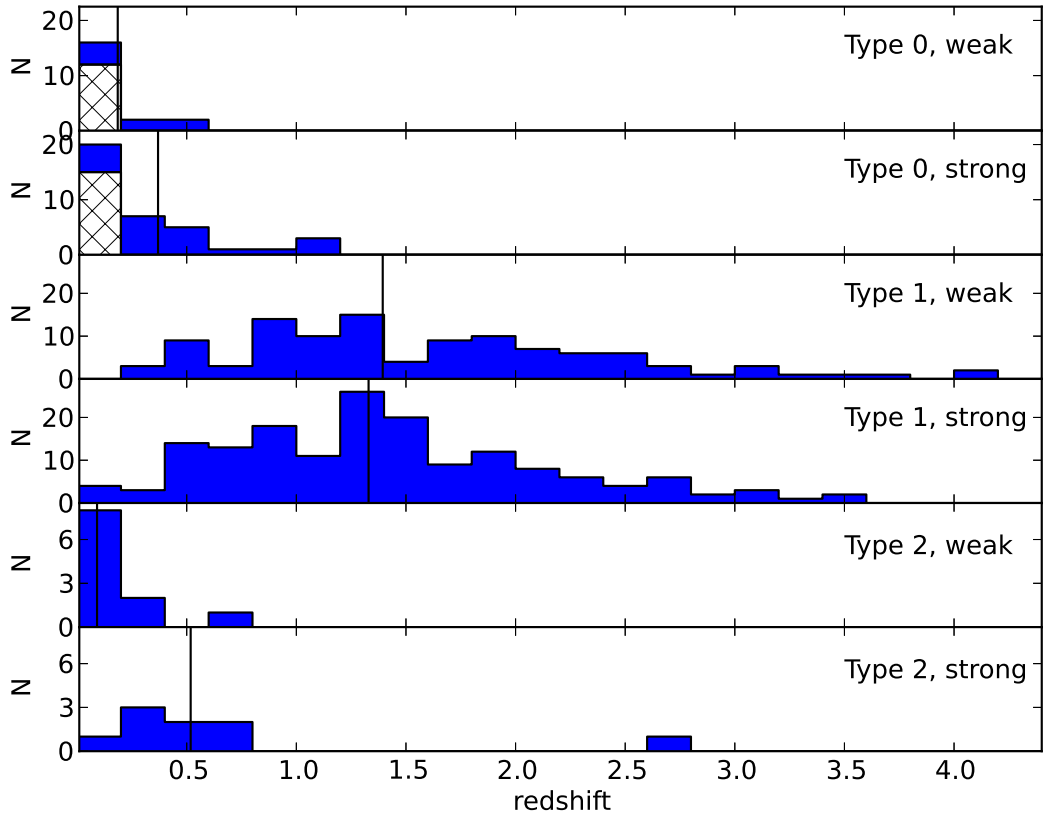}
\caption{
Redshift histograms of MASIV subsamples selected by 
5\,GHz flux density and  optical spectral type.
The BL Lacs with featureless spectrum are in the first
bin of Type 0 panels. The vertical line in 
each subplot indicates the median redshift.
}
\label{zHist}
\end{figure}

\clearpage
\begin{figure}
\epsscale{0.8}
\plotone{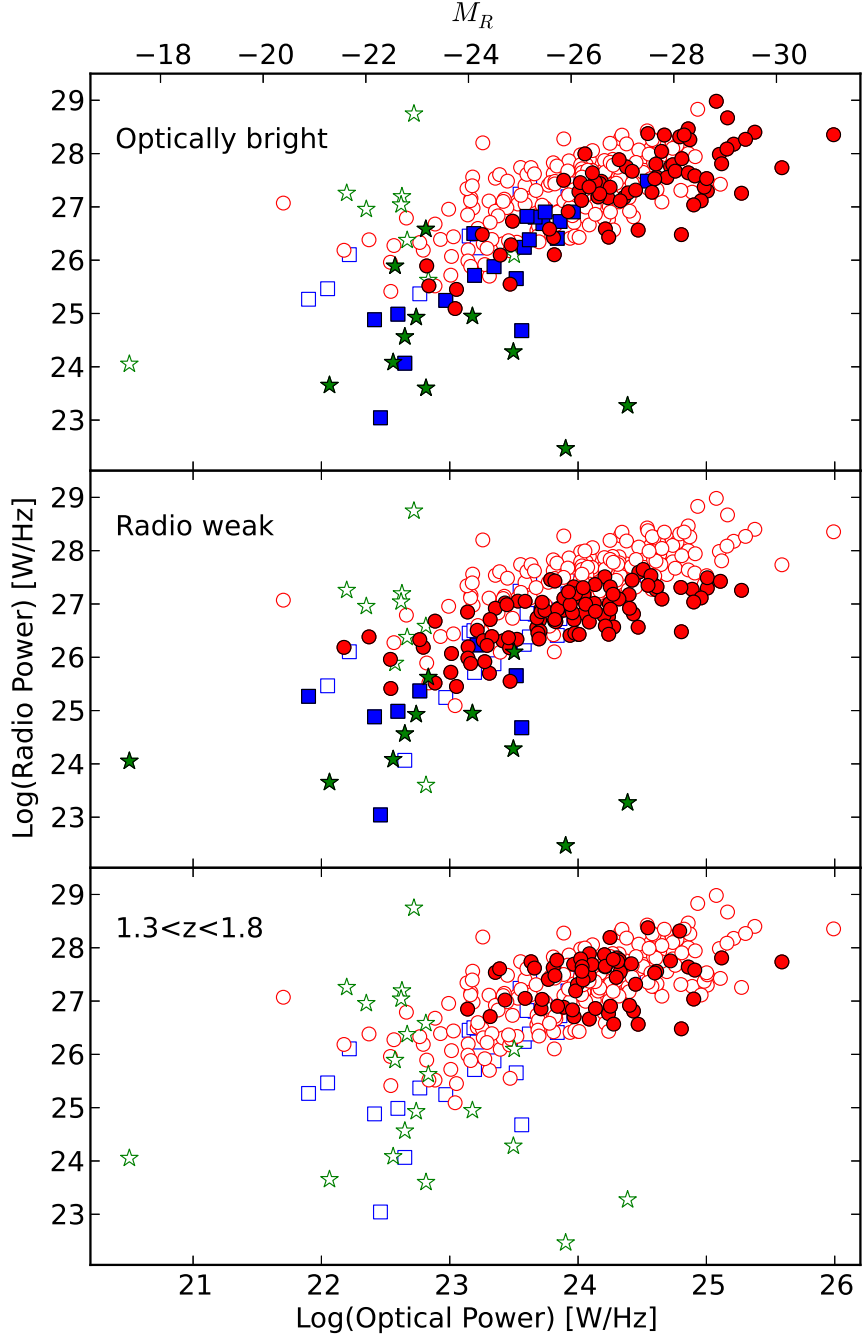}
%\plotone{opt_radio_lumin3e.eps}
\caption{ 
Optical R-band luminosity versus radio power, with FSRQ,
BL Lac and AGN marked with red circles,  blue squares and green stars
respectively. 
The filled symbols show in 
the top panel the location of optically bright sources, in the middle 
panel radio weak sources and in the bottom panel redshift 
$1.3<z<1.8$ objects.
Note the two  outliers (J0248+0434 and J1141+5953) 
near $M_R$=-26, whose magnitude estimates from the USNO-B1 
are too bright.
\label{rPowoLum}}
\end{figure}

\begin{figure}
\epsscale{1.0}
%\plotone{figD2dhistB.eps}
%\plotone{figD2hist.eps}
\plotone{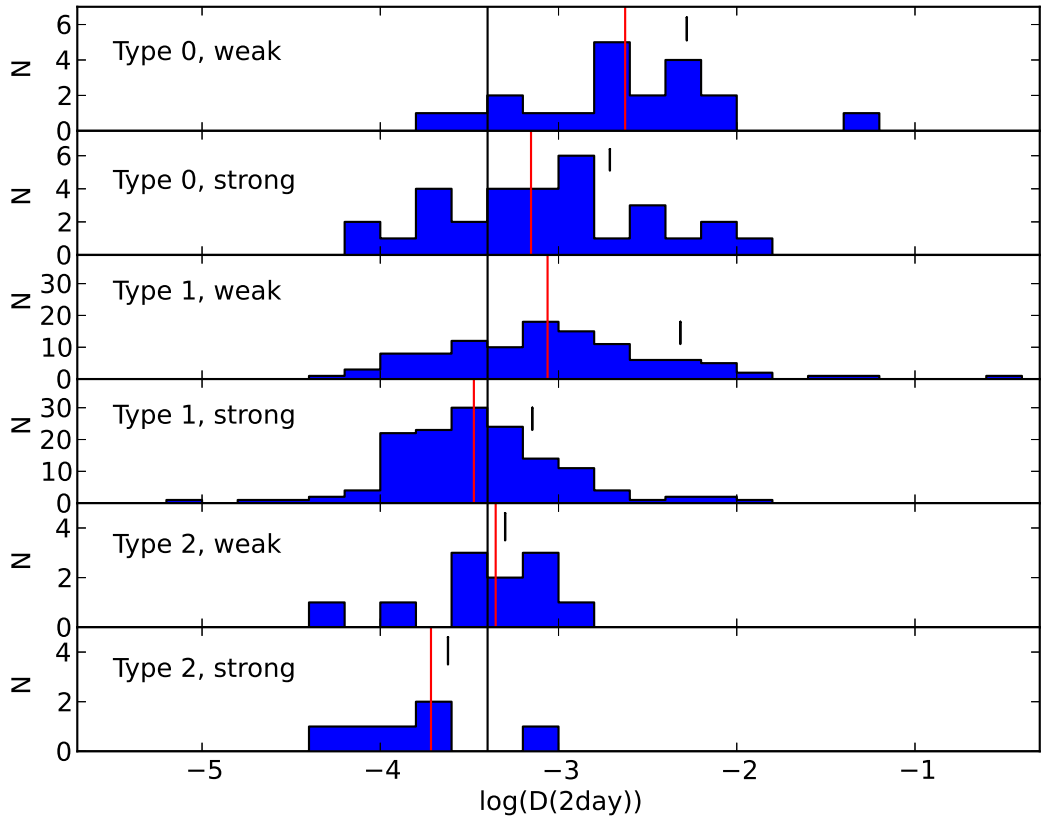}
\caption{
$D({\rm 2 days})$ histograms of MASIV subsamples selected by 
5\,GHz flux density and  optical spectral type.
The black vertical line at $-3.4$ indicates the limit where 
SF-analysis suggests variability. In each subplot 
the red vertical line indicates  the median value
and the black line the mean value.
}
\label{D2dHist}
\end{figure}

\begin{figure}
\epsscale{1.0}
%\plotone{Type1_z_vs_d2d_v2.eps}
\plotone{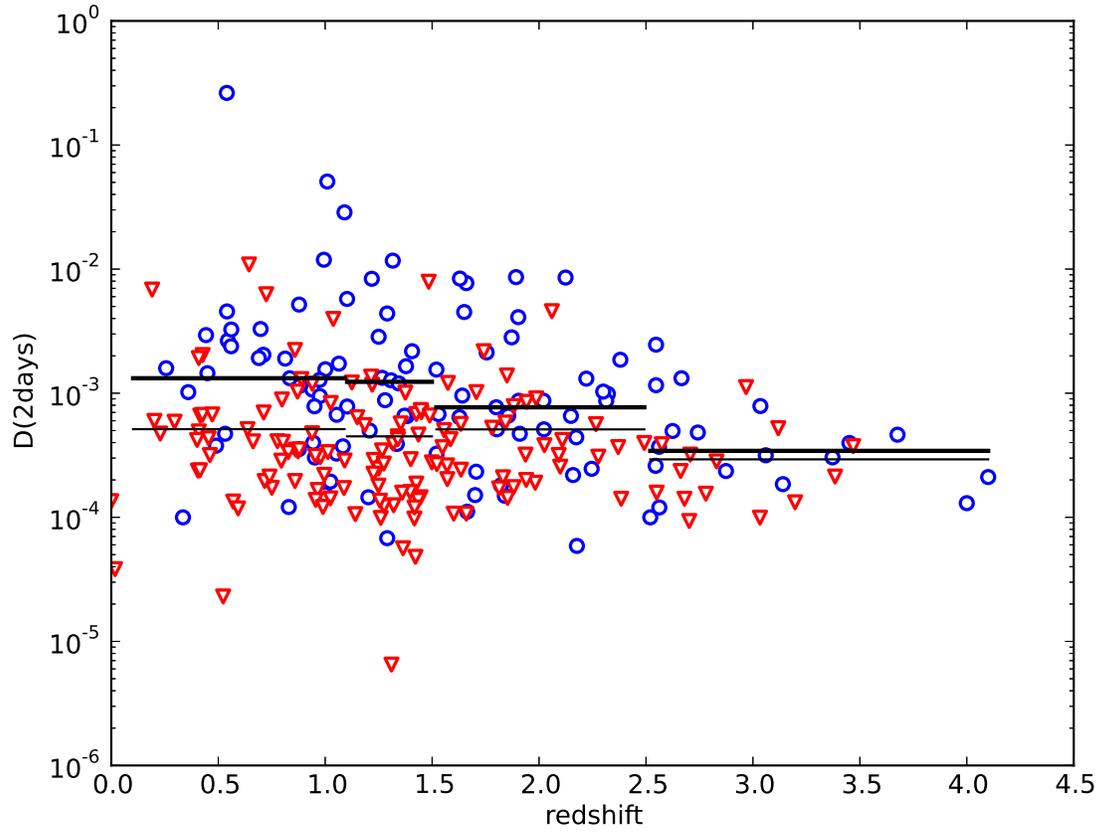}
\caption{
$D({\rm 2 days})$ against redshift for the Type 1 sources.
Red triangles indicate radio strong sources and
blue circles radio weak sources.
The horizontal thin and thick lines indicate the  
median $D({\rm 2 days})$ for the given redshift interval  
 for radio strong and weak samples,
respectively. Type 0 objects are not shown as most of 
them do not have redshifts. 
}
\label{D2dvsz}
\end{figure}

\clearpage

\begin{figure}
\epsscale{0.9}
\plotone{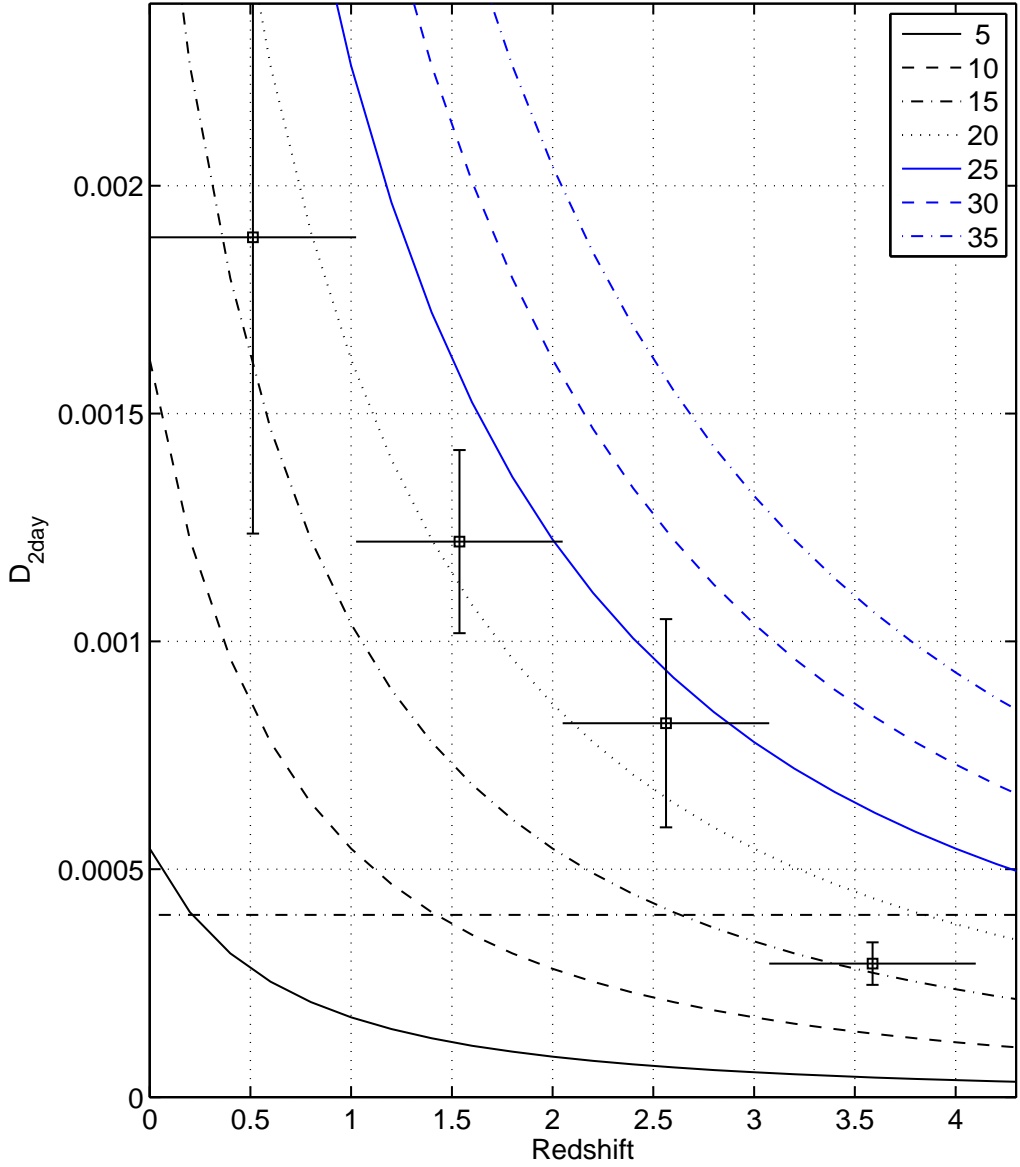}
%\plotone{D2d_zbin_theory.eps}
\caption{ 
The data for all FSRQs averaged into 4 redshift bins.  Over plotted are
theoretical curves for a simple ISS model, discussed in the text, 
with Doppler factors as indicated in the legend. 
The horizontal line is at 4
$\times 10^{-4}$, which is the threshold for significant variation.
}
\label{D_z}
%\end{figure*}
\end{figure}

\clearpage

\begin{deluxetable}{ccccc}
\tablecaption{
Summary of the optical identification using GSC2.3 
and SDSS DR5.
\label{ObjOptId}}
\tablewidth{0pt}
\tablehead{
\colhead{5$_{GHz}$limit} & \colhead{N} &\colhead{median} & \colhead{EF} &\colhead{survey}
}
\startdata
S$>$0.3Jy& 193& 18.2& 36&GSC2.3 $F$\\ 
S$<$0.3Jy& 167& 19.5& 79&GSC2.3 $F$\\ 
\cutinhead{$7^{h}<RA<16^{h}$ and $\delta <$ 64 }
S$>$0.3Jy& 62& 18.25&  4 & SDSS $r$\\ 
S$<$0.3Jy& 75& 19.82& 11 & SDSS $r$\\ 

\enddata
\tablecomments{Columns:(1) the radio flux density limit, 
(2) the number of detected optical counter parts
(3) the median magnitude of the sample (no galactic extinction correction)
(4) non-detections 
(5) The survey and pass band
}
\end{deluxetable}

\clearpage
\begin{deluxetable}{cccccccccclccc}
\tabletypesize{\scriptsize}
\rotate
\tablecaption{MASIV Data
\label{DataZMag}}
\tablewidth{0pt}
\tablehead{
\colhead{Source} & \colhead{RA} &\colhead{Dec}&\colhead{$S_{5GHz}$}  
&\colhead{Var} &\colhead{ $D({\rm 2 days})$}& \colhead{E(B-V)}& \colhead{R} &\colhead{Ref}
& \colhead{$z$} &\colhead{Type}& \colhead{ID} &\colhead{Ref}&\colhead{Comment}
\\
\colhead{(1)} &\colhead{(2)} &\colhead{(3)} &\colhead{(4)} &\colhead{(5)} &\colhead{(6)} &\colhead{(7)} &\colhead{(8)} &
\colhead{(9)} &\colhead{(10)} &\colhead{(11)} &\colhead{(12)} 
}
\startdata
J0005+3820& 00:05:57.1755& 38:20:15.168& 0.61& 1& 0.0897&0.08 &17.00& G& 0.229  & T1 &   fsrq &1   & \\
J0010+1724& 00:10:33.9920& 17:24:18.791& 0.84& 0& 0.0378&0.03 &16.96& G& 1.601  & T1 &   fsrq &2   & \\
J0017+5312& 00:17:51.7596& 53:12:19.125& 0.90& 2& 0.4006&0.18 &17.10& G& 2.574  & T1 &   fsrq &3,4 & \\
J0017+8135& 00:17:08.4750& 81:35:08.137& 0.61& 0& 0.1857&0.40 &15.49& G& 3.387  & T1 &   fsrq &5   & \\
J0019+2021& 00:19:37.8545& 20:21:45.570& 1.15& 1& 0.0604&0.06 &18.69& G&  ...   & T0 &   bll  &6,7 & \\
J0019+7327& 00:19:45.7871& 73:27:30.019& 0.44& 1& 0.3216&0.32 &17.48& G& 1.781  & T1 &   fsrq &8   & \\
\enddata
\tablecomments{Columns are as follows (1) IAU name (J2000.0), (2)R.A.(J2000.0), (3)Decl.(J2000.0)
(4) first epoch MASIV 5GHz flux density,
(5) The number of epochs of variability based on visual classification, where
'C' indicates secondary calibrators, (6) Structure function  $D({\rm 2 days})$, (7) Galactic extinction, 
(8) R-magnitude, (9) Reference; G=GSC2.3, S=SDSS DR5, U=USNO-B R2 
(10) Redshift, (11) Spectral type (12) Spectroscopic identification,
(13) Reference for redshift and spectroscopic identification, (14) Comment \\
References: 
(1) \citet{Stickel1994a}
(2) \citet{Wills1976},
(3) \citet{Sargent1989},
(4) \citet{Kuhr1983},
(5) \citet{Sowards-Emmerd2005},
(6) \citet{Chu1986},
(7) Palomar 2007 this paper,
(8) \citet{Lawrence1986}\\
Table 1 is published in its entirety in the electronic edition of <>, A 
portion is shown here for guidance regarding its form and content.
%references in table1_sort2.ref 
 }
\end{deluxetable}

\begin{deluxetable}{cccc}
\tablecaption{Objects with single emission line from our data
(that have an ambiguous identification). The range
indicates the useful wavelength range of the  spectrogram.
\label{ObjOneLine}}
\tablewidth{0pt}
\tablehead{
\colhead{Source} & \colhead{Line} &\colhead{Range} &\colhead{S/N$_{max}$}\\
&  \colhead{\AA} &\colhead{\AA} &\colhead{per pixel}
}
\startdata
J0213+3652&  4179&   3500-5000 &    \phn2 \\
J0342+3859 & 5453 &  4150-7500&     15      \\
J1316+6927&  5675&   4500-7500 &    \phn3 \\
J1711+6853&  5660&   3900-8500 &    \phn 8  \\
J1953+3537&  5125&   5200-7500 &    10\\
J2012+5308&  3967&   3950-7500 &    \phn 6\\
\enddata
\end{deluxetable}

\begin{deluxetable}{clcc}
\tablecaption{
Summary of the spectroscopic identification. 
The third and fourth column indicates the number of sources
in the radio strong and radio weak samples, respectively.
\label{SpecIdSumm} 
}
\tablewidth{0pt}
\tablehead{
\colhead{Type} & \colhead{Spect id} &\colhead{Strong} 
&\colhead{Weak} 
}
\startdata
Type 0 & BLL &   \phn37 & \phn11 \\
       & BLL?  & &\phn\phn 5  \\
       & PEG &  &\phn \phn6 \\
Type 1 & FSRQ & 155 & 107 \\
       & FSRQ? & \phn\phn1 &  \\
       & Sy 1 & \phn\phn & \phn\phn2  \\
       & Sy 1/BLRG & \phn\phn1 &   \\
       & NLSy1& \phn\phn1 &   \\
       & NLQSO & \phn\phn2 & \\
       & Sy1.5/LINER &\phn\phn 1 & \\
Type 2 & Sy2 & \phn\phn3 &\phn\phn 2 \\
       & Sy2? & & \phn\phn1\\
       &NLRG & \phn\phn5 &\phn\phn 1 \\
       & LINER  & \phn\phn1 &\phn\phn 1  \\
       & HII? &  &  \phn\phn1\\
       & galaxy  &  &\phn \phn3 \\
Single line objects &  &\phn\phn 1 & \phn\phn5 \\

\enddata
\end{deluxetable}

\begin{deluxetable}{crrrr}
\tablecaption{
Classification break down of the radio strong and weak samples
including the  secondary calibrators. The 'SDSS sample' 
source distributions are in column four and five. 
\label{StypeTab}}
\tablewidth{0pt}
\tablehead{
\colhead{type } & \colhead{N$_{full}$} &\colhead{\%} &\colhead{N$_{SDSS}$}
&\colhead{\%} 
}
\startdata
\cutinhead{Radio strong sources}
Type 0 & 37 & 18$\pm$3\%&19&24$\pm$6 \%\\
Type 1& 162 & 78$\pm$6\%&57&72$\pm$10\% \\ 
Type 2&   9 &  4$\pm$1\%& 2& 3$\pm$2 \%\\  
\cutinhead{Radio weak sample}
Type 0 & 22 &16$\pm$3\% & 8&14$\pm$5\%\\   
Type 1& 108 & 78$\pm$7\%& 46& 78$\pm$11\%\\  
Type 2&   9 &6$\pm$2\%  &   5&8$\pm$4\%\\  
\enddata
\tablecomments{
The errors are Poissonian.
}
\end{deluxetable}

\begin{deluxetable}{lcccccl}
\tablecaption{Summary of some flat radio spectrum AGN surveys.
Note, the PEGs in CBS and MASIV are included in the ``other AGN'' as
is the case with CGRaBS and DXRBS.
\label{BllFsrqOther}}
\tablewidth{0pt}
\tablehead{
\colhead{Survey}& \colhead{5\,GHz limit} & \colhead{R-mag}& \colhead{FSRQ} &\colhead{BL Lac } &\colhead{other AGN}
&\colhead{Comment}\\
&\colhead{$mJy$}& & \colhead{\%}& \colhead{ \%}& \colhead{ \%}
}
\startdata
CBS   &30  &17.5&45   & 21 & 34 &   \\
CBS rb &65: &17.5&51   & 20 & 29& MASIV 5\,GHz limit\\   
CGRaBS  &65 &23.7&88   & 10 & 2& Selection by $Figure\, of\, Merit$\\
DXRBS   & 50 &24.0&80   & 12 & 8& some steep spectrum sources \\
MASIV   &60  &23.1&78   & 15 &  7&  Selection by 8.4\,GHz \\
MASIV ob  &60  &17.5:&64   & 24 &  12& Optically bright CBS limit \\
MASIV ISS& 60&21.3& 64  & 34 &  2& Persistent IDV sources, IDV3, 4\\
MASIV SDSS& 60&21.7& 75  & 18 &  7& SDSS sample

\enddata
\tablecomments{Columns:(1) the name of the sample, (2) 5\,GHz 
radio flux density limit for faintest objects, 
(3) optical magnitude limit or the faintest detected objects, 
(4) the fraction of broad line objects, 
(5) the fraction of BL Lac objects
(6) the fraction of other spectral type objects, mostly narrow line objects.\\
The colon indicates MASIV radio flux density limit for the ``CBS rb", 
and CBS optical limit for ``MASIV ob''.
Note, the DXRBS includes steep radio spectrum sources as well
\citep{Landt2001}.
}
\end{deluxetable}

\begin{deluxetable}{ccrcr}
\tablecolumns{5} 
\tablecaption{Median $D({\rm 2 days})$ 
of the subsamples selected by optical
spectral type and 5\,GHz flux density. The First column indicates 
optical spectral type, followed by the median $D({\rm 2 days})$
and the number of objects.
}
\tablewidth{0pt}
\tablehead{
\colhead{Type} &\colhead{$D({\rm 2 days})\times$1000} &\colhead{N}
& \colhead{$D({\rm 2 days})\times$1000} &\colhead{N}  \\
&  \colhead{$S_{5GHz}>0.3Jy$}& &\colhead{$S_{5GHz}<0.3Jy$}
}
\startdata
0    & 0.701 & 31 \ & 2.35 & 22\\
1    & 0.335&143 \ & 0.868 &108\\
2    & 0.193&  6 \  & 0.44 &  9\\
\cutinhead{Type 1 $z<1.4$ sources}
1 & 0.346 & 79 & 1.27 & 54\\
\enddata
\label{D2dTable}
\end{deluxetable}

\begin{deluxetable}{lrrrrrr}
\tablecaption{
Number of sources and the fractions of ``variable'' MASIV epochs
for the three optical spectral types.
The first column indicates the number variable in epochs and
the next columns show the
fraction of sources in each group with  the number of sources in 
parentheses.
\label{VarEpochs}}
\tablewidth{0pt}
\tablehead{
\colhead{Sample} &\multicolumn{2}{c}{Type 0}&\multicolumn{2}{c}{Type 1}&
\multicolumn{2}{c}{Type 2}\\
\colhead{IDV} &\colhead{strong}&\colhead{weak} &\colhead{strong}&\colhead{weak} &\colhead{strong}&\colhead{weak}
}
\startdata
0  &13\% (4)  &18\% \, (4)  &36\% (51) &36\%(39)&5&7\\
%0c & (2)      &(0)       &(13)     &(0 )&2&0\\
1  &19\% (6)  &18\% \, (4)  &24\% (34) &15\%(16)&1&1 \\
%1c & (2)      &(0)      &(3)     &(0)&1&0 \\
2  &13\% (4)  &(0)      &24\% (34) &17\%(18)&0&0 \\
%2c & (2)      &(0)      &(3)     &(0)&0&0 \\
3  &29\% (9)  &18\% \, (4)  &8\% (12)  &18\%(19)&0&1 \\
4  &26\% (8)  &46\% (10) &8\% (12)  &15\%(16)&0&1  \\
\enddata
\end{deluxetable}

\begin{deluxetable}{lrcccrccc}
%\tabletypesize{\scriptsize}
%\rotate
\tablecaption{Redshift, core radio power and optical luminosity,
excluding secondary calibrators. The IDV0, IDV1, etc. indicates the number 
of epochs of variability \citep{Lovell2008}. 
\label{RadioLuminT}}
\tablewidth{0pt}
\tablehead{
&\multicolumn{4}{c}{Strong sources}  &\multicolumn{4}{c}{Weak sources} \\
\colhead{Sample}& \colhead{N } & \colhead{$<z>$ }  & \colhead{log($P_r$)} & \colhead{$M_R$}& \colhead{N } & \colhead{$<z>$ }  &\colhead{log($P_r$)}& \colhead{$M_R$}\\
&&&\colhead{$W m^{-2}$}&&&&\colhead{$W m^{-2}$}
}
\startdata
Type 0 & 16&0.4$\pm$0.3 &26.6$\pm$0.6 & -25.2$\pm$1.1&  10&0.2$\pm$0.1& 25.0$\pm$0.8& -23.6$\pm$1.3\\ 
Type 1 &142&1.3$\pm$0.7 &27.6$\pm$0.6& -26.4$\pm$1.6& 108&1.4$\pm$0.8& 26.8$\pm$0.5& -25.9$\pm$1.6 \\ 
Type 2 &  6&0.4$\pm$0.3 &26.7$\pm$1.3 & \nodata          &  9&0.1$\pm$0.2&24.1$\pm$1.1 &\nodata  \\
\hline
&\multicolumn{7}{c}{Type 1}\\
$D({\rm 2 days})$$>4\times10^{-4}$&59&1.2$\pm$0.6&27.5$\pm$0.6&-26.2$\pm$1.4&76&1.3$\pm$0.7&26.7$\pm$0.5&-25.7$\pm$1.5\\
$D({\rm 2 days})$$<4\times10^{-4}$&83&1.4$\pm$0.7&27.7$\pm$0.5&-26.7$\pm$1.6&32&1.8$\pm$1.0&27.0$\pm$0.5&-26.7$\pm$1.4\\
&\multicolumn{7}{c}{ 1.0 $<z<$ 1.8}\\
$D({\rm 2 days})$$>4\times10^{-4}$&24&1.4$\pm$0.2&27.5$\pm$0.2&-26.4$\pm$1.0&29&1.3$\pm$0.2&26.7$\pm$0.2&-26.0$\pm$1.2\\
$D({\rm 2 days})$$<4\times10^{-4}$&35&1.4$\pm$0.2&27.6$\pm$0.2&-26.4$\pm$1.3&10&1.3$\pm$0.3&26.8$\pm$0.2&-25.7$\pm$1.0\\
\hline
&\multicolumn{7}{c}{Type 1}\\
ISS (IDV234)&58& 1.2$\pm$0.7&27.5$\pm$0.6&-26.3$\pm$1.4&53&1.2$\pm$0.6&26.7$\pm$0.5&-25.4$\pm$1.4\\
IDV0        &50& 1.4$\pm$0.8&27.7$\pm$0.6&-26.7$\pm$1.8&39&1.9$\pm$1.0&27.1$\pm$0.5&-26.5$\pm$1.6\\
IDV1        &34& 1.3$\pm$0.6&27.5$\pm$0.5&-26.3$\pm$1.3&16&1.4$\pm$0.7&26.8$\pm$0.4&-26.0$\pm$1.3\\
IDV2        &34& 1.5$\pm$0.7&27.6$\pm$0.6&-26.4$\pm$1.5&18&1.3$\pm$0.7&26.7$\pm$0.5&-26.0$\pm$1.3\\
IDV3        &12& 1.0$\pm$0.5&27.4$\pm$0.6&-25.9$\pm$1.2&19&1.1$\pm$0.7&26.6$\pm$0.6&-25.0$\pm$1.4\\
IDV4        &12& 1.4$\pm$0.7&27.6$\pm$0.4&-26.5$\pm$1.3&16&1.2$\pm$0.5&26.7$\pm$0.4&-25.3$\pm$1.4  \\
\enddata
\tablecomments{Columns: (1) the sub-sample, (2) Number of objects, (3) Median redshift, 
(4) Median of the Log radio power with standard deviation, (5) Median absolute magnitude with standard deviation,
(6)-(9) same as Cols. (2)-(5).
}
\end{deluxetable}

\appendix
\section{COMMENTS ON ARCHIVAL REDSHIFTS}

After careful inspection three objects with archival redshift were rejected.
In addition, unreliable redshifts have been proposed for ten BL Lac objects
and four objects with adopted, but uncertain redshift or 
spectroscopic identification.
We briefly discuss these here. 

\noindent
J0200+0322:  \citet{Fricke1983} reported redshift,  $z$=0.765,
based on a  single (\ion{Mg}{2}$\lambda$2798) line at 4920\AA.
Their spectrum had the wavelength range from 4200-7100\AA ~\ where
for a typical QSO spectrum a single broad line can be 
detected at $z \sim$ 0.8 (\ion{Mg}{2})
or 1.6 (\ion{C}{3}]).\\
\noindent
J0204+1514:  \citet{Stickel1996}  reported $z$=0.833 based on two
narrow emission lines. Later \citet{Perlman1998} found more lines and a 
better fit for $z$=0.405 and
they classified this object as an FSRQ, which we  adopted for this work. 
However this object should be re-observed in order to secure 
the spectroscopic identification.\\
\noindent
J0406+2511: \citet{Hook1996} assumed the single line at 6260\AA ~\ 
to be  [\ion{O}{2}]$\lambda$3727,  at $z$=0.68.
Their data has a wavelength range from 5200-9100\AA, however the red and blue 
ends are very noisy, reducing the useful range to   5800-8500\AA.
If the redshift is indeed 0.68, H$\beta$ should be around 8150\AA,
and/or   [\ion{O}{3}]$\lambda$5007  around 8410\AA,
which are not seen. \\
\noindent
J0449+1121: \citet{vonMontigny1995} reported $z$=1.27, 
however \citet{Halpern2003} 
found featureless spectrum. Thus we adopted a BL Lac identification 
for this source.\\
\noindent
J0738+1742: NED reports  $z=$ 0.424. However, this is the 
redshift of intervening 
MgII absorption feature, i.e. $z \gtrsim$ 0.43 (\citet{Rector2001} 
and references therein). 
\\ 
\noindent
J0745+1011: We adopted the redshift from \citet{Best2003} who
 measured $z$=2.624, however, \citet{Labiano2007} could not 
confirm this. If the redshift 
 is correct the radio power is 4$\times$10$^{28}$W/Hz and an  outlier in the 
 $M_R$-Log($P$)-plot (Figure~\ref{rPowoLum}). 
This object should be re-observed.\\
\noindent
J0753+5352: This source has featureless optical spectrum 
and the redshift has a lower limit of 0.2  based on imaging data
\citep{Stickel1993a}. \\
\noindent
J0818+4222: This is a radio source with featureless optical spectrum and
has no reliable redshift available 
(\citet{Rector2001} and references therein).\\
J0929+5013: The SDSS redshift is uncertain as there are no obvious emission 
lines. \\
\noindent
J1150+2417: This object has a featureless optical spectrum 
\citet[][and references therein]{Rector2001}.\\
\noindent
J1309+1154  This object has a featureless optical spectrum by
\citet{Lynds1972} and SDSS. Note that this object has 
SDSS DR5 redshift of 0.3852
and SDSS DR 8 redshift of 2.601. 
\\
\noindent
J1446+1721: The redshift of this FSRQ is uncertain 
as \citet{Sowards-Emmerd2005} reported, $z$=0.102, 
however \citet{Healey2008} reported $z$=1.026. For this work 
we adopt FSRQ classification with unknown redshift.\\
\noindent
J1502+3350: \citet{White2000} report featureless spectrum for this object.
Note that the SDSS DR6 suggest $z$=2.222, however the spectrum appears 
featureless.\\  
\noindent
J1602+3326: The redshift is based on optical magnitude \citep{Snellen2000} 
using the GPS galaxy Hubble
diagram \citep{Stanghellini2005}.\\
\noindent
J1648+2141 The SDSS spectrum is featureless, however the SDSS DR6 
suggest $z$=1.085.\\
\noindent
J1649+7442: This redshift is from \citet{Appenzeller1998}
from the spectrum with  a wavelength range  from 4000 to 7000/9000\AA.
The redshift is based on a  single line at 6658\AA ~\ which is 
assumed to be \ion{Mg}{2}$\lambda$2798 at $z$=1.378.
At this redshift, one would expect to see \ion{C}{3}]$\lambda$1909 at
4532\AA, however if the line was \ion{C}{3}] or H$\beta$
one would expect to see  \ion{C}{4} at $\sim$5392\AA ~\ or 
[\ion{O}{3}] near 6870\AA.
The redshift is adopted, but  needs to be confirmed.\\
\noindent
J1719+1745:  \citet{Sowards-Emmerd2005}, reported $z$=0.173. However 
\citet{Shaw2009} find featureless spectrum 
and they estimate redshift to be $>$0.58. This source is classified as a 
BL Lac object without known redshift.\\
\noindent

\section{TRANSFORMATION FROM GSC2.3 TO LANDOLT SYSTEM}

In order to transform the GSF2.3 magnitudes to the 
\citet{Landolt2009} system,  we obtained $JFN$ magnitudes 
from  the Landolt 
standard stars. We included only stars with four or more
measurements by Landolt, then plotted
$J-B$, $F-R$ and $N-I$  and removed outliers. 
Least squares polynomial fit resulted 
 following  transformations:
\begin{equation}
B_{GSC}=J+0.18(J-F)-0.08, 
\end{equation}

\begin{equation}
R_{GSC}=F[-0.085(F-N)+0.04],
\end{equation}

\begin{equation}
I_{GSC}=N[+0.066(F-N)-0.02].
\end{equation}

The terms in square brackets for $R_{GSC}$ and $I_{GSC}$
are from the best least-squares fit, 
however the fit is improved only marginally, 
hence we have omitted the color term.
The root mean squared error for the 
BRI-conversions are
0.10, 0.11 and 0.11, respectively.

The  $B_{GSC}$ transformation is valid for the $(J-F)$-colors
from -0.5 to 2.5 and 
$R_{GSC}$,  $I_{GSC}$ for the $(F-N)$-color range from  -0.4 to 1.4.

\section{ NOTES ON INDIVIDUAL OBJECTS}

In this appendix, we discuss some of 
our spectroscopic follow-up targets.
For the narrow line objects we report the 
EW of the strongest line.

\noindent
J0019+2021 BL Lac classification is from \citet{Chu1986} who found no strong
emission lines. Our DBSP data (S/N$\lesssim$35) 
shows absorption system at 
$z$=1.41 with \ion{Mg}{1}$\lambda$2852, 
\ion{Mg}{2}$\lambda$$\lambda$2803,2796, (\ion{Fe}{2}$\lambda$2344)
and a possible absorption system at $z$=2.274 (Figure~\ref{ExamSpec}). \\
J0150+2646 The DBSP (07-2007) and NOT (8-2004) spectrum are featureless with 
 S/N$\lesssim$14 and S/N$\lesssim$8, respectively. Thus, this source is 
tentatively identified as a BL Lac object.  \\
J0509+0541 \citet{Halpern2003} reported featureless spectrum from several
observations. Our high S/N ($>$130) data  is featureless showing only galactic
absorption lines and unidentified absorption at 4276$\AA$.\\
J0800+4854 Our data show strong absorption 
at the \ion{C}{4}$\lambda$1549- and Ly$\alpha$-lines
and also  the \ion{He}{2}$\lambda$1640-line is easily detected. \\
J1118+2922 This low redshift object has two easily detected narrow 
emission
lines at $\lambda$3998\AA ~\  and $\lambda$7038\AA, 
identified as  [{\ion{O}{2}]$\lambda$3727 and
H$\alpha$ at redshift of 0.072. The line ratios suggest HII region or
 Liner classification. \\
J1352+3603  We detect  narrow  [\ion{O}{3}]$\lambda$$\lambda$4958, 5007 
and  [\ion{O}{2}]$\lambda$3727  lines 
at $z$=0.3057 with, 
M$_B$=-21.5, (M$_R$=-23.0),  C=42\%,  
EW([\ion{O}{2}])=20\AA  ~\   and  
noisy EW(H$\alpha$)$\sim$ 37\AA.
The source is classified as PEG based on the Ca-break and line EW. 
On the other hand, based on the classification scheme of \citet{Landt2004}, 
this is a strong-lined source. \\
J1442+0625 Our data show a single line at 4753\AA, with continuum from 
4800-7500\AA ~\ (S/N$\lesssim$10). Our  tentative identification is 
\ion{Mg}{2}$\lambda$2798-line at $z$=0.698.
If this line would be \ion{C}{3}]  at $z$=1.49 we would expect to see 
\ion{Mg}{2}$\lambda$2798 at about 7000\AA~\ which is not detected.\\
J1444+0257  This object has a featureless spectrum 
from 4500 to 7500\AA ~\ with S/N$\lesssim$5,
possibly a BL Lac object\\
J1444+1632 We detect a  single line at 5688\AA ~\ with 
noisy continuum (S/N $\lesssim$3) and  tentative
line at 7247\AA, these are  
identified as Ly$\alpha$ and \ion{C}{4}$\lambda$1549,
which would indicate $z$=3.7.
\\
J1505+0326 
 This object has been classified as FSRQ in CGRaBS \citep{Healey2008}.
Our NOT data confirms the CGRaBS and SDSS redshift.
However reanalyzing the SDSS data and based on the 
NOT data the rest frame line widths are less than  2000km/s
indicating NLSy1 classification.\\
J1651+0129 Our DBSP data shows a possible line at 7128\AA ~\
with the continuum S/N$\lesssim$7. As we are not able to identify the line
or spectrum, this target is not included to the analysis.\\
J1718+4448  The  DBSP red spectrum show no obvious 
lines from 5300 to 8300\AA~\ with S/N $\lesssim$ 5,
possibly a BL Lac object.\\
J1728+1931 We detect  narrow  ([\ion{O}{2}]$\lambda$3727), 
 [\ion{O}{3}]$\lambda$$\lambda$4958, 5007  
and a noisy  [\ion{O}{1}]$\lambda$6300 lines at $z$=0.1759,
with   [\ion{O}{3}]$\lambda$5007/H$\beta$$>$3,
 EW([\ion{O}{2}])=13\AA,
 M$_B$=-21.1, (M$_R$=-24.2) and   C=42\%. 
The H$\alpha$-line is noisy, however it appears  to be narrow. 
This is border line object between strong and weak lined sources using 
\citet{Landt2004} classification.
We  tentatively classify this as a  Sy2-type object.
\\
J1734+3857 \citet{Stickel1989}
classification for this source is  
an FSRQ  at $z$=0.97 based on a single 
broad  emission line which was assumed to be 
\ion{Mg}{2}$\lambda$2798  (EW(\ion{Mg}{2})=15.3\AA, observed)
Our data shows also only a 
single line at 5535\AA  ~\ assumed to be \ion{Mg}{2} at 
$z$=0.97 with the EW=4.2\AA, 
hence BL Lac classification. \\  
J1739+4737 This is a  BL Lac-object  with featureless spectrum 
\citep{Scarpa2000}. Our DBSP data show
 no lines with S/N$\lesssim$20\\
J1742+5945 
This object has a featureless spectrum 
\citet{Laurent-Muehleisen1998},  
and the redshift ($z$=0.4) is based on a detection of the
host galaxy \citep{Nilsson2003}. Our 
NOT spectrum (07-2006) shows  no lines from 
4000 to 7500 \AA ~\ with S/N$\lesssim$25.  \\  
J1747+4658 
This source has no emission line redshift. 
\citet{Vermeulen1996}  reported
an intervening \ion{Mg}{2} %$\lambda$  
absorption line 
system at $z$=1.484, which, 
by some authors, has been adopted as 
the source redshift.
Our NOT spectrum (Grism 6) from  4000 to 5500\AA ~\ 
with  S/N$<$6 is featureless.\\
J1757+0531 
Narrow emission lines,  [\ion{O}{2}]$\lambda$3727 
and   [\ion{O}{3}]$\lambda$$\lambda$4958, 5007 
at $z$=0.34482 
are clearly visible and so is a noisy strong H$\alpha$ feature at the red 
end of the spectrogram. 
The line and continuum properties suggest Type 2 identification:
[\ion{O}{3}]$\lambda$5007/H$\beta$$>$3, ~\ M$_R$=-23.2, 
~\ EW([\ion{O}{3}]$\lambda$5007)=37\AA ~\  and  
  C=28\%.  However the  continuum is rather noisy,
(S/N$\lesssim$8) and weak broad emission lines might have been undetected
(Figure~\ref{ExamSpec}).  
Tentative spectroscopic identification is NLRG,
however, based on Fig. 8 and 9 of \citet{Caccianiga2008}
this object could be a Type 1 object with the AGN core diluted by 
the host galaxy.\\
J1832+1357 Our data confirms the CGRaBS \citep{Healey2008} redshift.
We detect also some absorption features in \ion{C}{4}$\lambda$1549-line\\
J1848+3219 \citet{Sowards-Emmerd2005} reported  this source as an 
FSRQ at  $z$=0.798.
Our data shows a strong line at 5035\AA, a noisy narrow line
at 6703\AA ~\ (identified as \ion{Mg}{2}$\lambda$2798 and 
  [\ion{O}{2}]$\lambda$3727, respectively) 
and   no strong lines in 
3600-5000 and 5100-7500\AA ~\ with  S/N$\lesssim$17, supporting
  the literature identification and redshift.\\
J1905+1943 A strong broad line is detected at 5106\AA \, 
and a tentative broad line
at 6267\AA, suggesting $z=$2.3.\\  
J1950+0807 From the DBSP data we detect 
narrow  [\ion{O}{2}]$\lambda$3727,
   [\ion{O}{3}]$\lambda$$\lambda$4958, 5007  and 
 broad H$\alpha$-lines  at $z$=0.2964, with 
 M$_R$=-22.8,  M$_B$=-20.7 and EW([\ion{O}{3}]$\lambda$5007)=-48\AA, 
suggesting Sy1.9 classification.\\
J2023+5427 The redshift 1.48 is measured  from  \ion{Mg}{2}$\lambda$2798
and \ion{C}{3}]$\lambda$1909-lines. 
In addition we detect  MgI(2852), MgII(2803,2796), FeII(2387,2600) and 
FeII(2383,2344) absorption lines at $z$=1.415\\
J2130+0339 From the DBSP data we detect  narrow 
lines  [\ion{O}{2}]$\lambda$3727 
and   [\ion{O}{3}]$\lambda$$\lambda$4958, 5007 at $z$=0.653, 
with  
EW([\ion{O}{3}]$\lambda$5007)=140\AA,  M$_R >$-22.5 and C=31\%,
suggesting Type 2 identification.
Note, this source is  not detected in the GSC2.3. \\ 
J2155+0916   This object is 
tentatively identified as a BL Lac object. Our NOT spectrum is
featureless  from 4500 to 7500\AA ~\ with  S/N$\lesssim$20,
however if the redshift is  $\sim$0.6, both  
\ion{Mg}{2}$\lambda$2798 and H$\beta$-lines will be outside
the spectrogram.\\
J2201+5048 We confirm the \citet{Sowards-Emmerd2005} redshift, 1.899 and 
in addition we detect \ion{Mg}{2}-absorption system at $z$=1.3822
 with \ion{Mg}{1}$\lambda$2852, \ion{Mg}{2}$\lambda$$\lambda$2803,2796, 
\ion{Fe}{2}$\lambda$2344,2374,2383, however this source  was
rejected from the \citet{Lovell2008} analysis due to evidence of source structure, 
hence not included to the  analysis.\\
J2203+1725  \citet{Smith1977} classified this source as a FSRQ 
based on strong \ion{C}{3}]$\lambda$1909- 
and \ion{Mg}{2}$\lambda$2798-lines  at $z$=1.076.
Our data shows a single line at 
5811\AA ~\ with  EW(\ion{Mg}{2})=3.3\AA ~\ and S/N$\lesssim$60
around the  line, hence  BL Lac object classification.\\
J2212+2759 This is a  BL Lac object, with no detected emission lines,  
based on our DBSP data (07-2007) 
with S/N$\lesssim$45 and  NOT data (07-2005)
with S/N$\lesssim$18.\\  
J2230+6946 We confirm the \cite{Healey2008} (CGRaBS) redshift.\\
J2237+4216 An emission line is detected at 5655\AA \, and this is the only 
emission line between 4500 and 7500\AA, suggesting \ion{Mg}{2}  
identification.\\
J2241+4120  \citet{Henstock1997} and 
\citet{Sowards-Emmerd2005} 
classified this as a BL Lac object. 
Our DBSP data (07-2007) with  S/N$\lesssim$60 
and NOT data (07-2005) from 4000 to 7500\AA ~\ with 
S/N$\lesssim$30 confirm this.\\ 
J2242+2955 A strong broad line is detected at 4102\AA \, 
and a tentative broad line
at 5013\AA, suggesting $z=$1.7.\\  
J2258+0516 We detected a strong absorption feature at the centre of the 
\ion{C}{4}$\lambda$1549-line and Ly$\alpha$ line
is almost fully absorbed.\\
J2303+1431 This is a BL Lac object with featureless spectrum.
Our DBSP spectrum  (07-2007)  has
 S/N$\lesssim$25 and NOT spectrum (07-2006)  S/N$\lesssim$8. \\
J2315+8631 A broad line at 5273\AA ~\ (\ion{Mg}{2}$\lambda$2798) is visible
in our DBSP (08-2007)  and NOT (07-2006) spectrum.
This is the only strong emission line between 3700 and 8500\AA ~\
(S/N$\lesssim$30) ,
hence a tentative redshift of $z$=0.88.\\
J2325+3957 \citet{Shaw2009} reported a featureless 
spectrum with 
intervening \ion{Mg}{2}$\lambda$2798 and FeII  
absorption features  at $z$=1.04.
Our DBSP data (07-2007) shows no emission lines. However, we detect
absorption features which are identified as 
\ion{Mg}{1}$\lambda$2852, \ion{Mg}{2}$\lambda\lambda$2803,2796, 
\ion{Fe}{2}$\lambda$2600 (and \ion{Fe}{2}$\lambda$2344)
at $z$=0.4695  
and \ion{Mg}{2}$\lambda\lambda$2803,2796, 
\ion{Fe}{2}$\lambda$2600 
at  $z$=0.4156 with the  continuum  S/N
up to  $\sim$25.
We also detected an absorption feature at 5757\AA  ~\ possibly
\ion{Mg}{2}$\lambda$2798 at $z$=1.04, 
but not the expected \ion{Fe}{2} feature at 5353\AA.
The 4119\AA ~\ feature,  \ion{Mg}{2}$\lambda$2803 at $z$=0.4695, 
 is also visible from the NOT spectrum (07-2006).\\

\end{document}